\documentclass[preprint2]{aastex} 
\usepackage{graphicx}
\usepackage{natbib}
\usepackage{blindtext,lineno,ulem}
\usepackage{times}
\usepackage{longtable}
\usepackage{amsmath}
\usepackage{textcomp}
\usepackage{amssymb}
\usepackage{lscape}
\usepackage{xcolor}

\usepackage[caption=false]{subfig}

\shortauthors{Kaur et al.}

\def\asec{\ifmmode^{\prime\prime}\else$^{\prime\prime}$\fi}
\def\farcs{\hbox{$ \,\!\!^{\prime\prime}$}}       
\def\degs{\ifmmode ^{\circ}\else$^{\circ}$\fi}

\begin{document}
\title{Identifying the 3FHL catalog: I. \\
Archival $Swift$ Observations and Source Classification}
\author{A. Kaur\altaffilmark{1},  M. Ajello\altaffilmark{2},  S. Marchesi\altaffilmark{2},N. Omodei\altaffilmark{3}
\affil{\altaffilmark{1}Department of Astronomy and Astrophysics, 525 Davey Lab, Pennsylvania State University,
University Park, PA 16802, USA}
	}
\affil{\altaffilmark{2}Department of Physics and Astronomy, Clemson University, SC 29634-0978, U.S.A.}
\affil{\altaffilmark{3}Department of Physics and SLAC National Accelerator Laboratory, Stanford University, Stanford, CA 94305, USA}
\begin{abstract}
We present the results of an identification campaign of unassociated sources from the $Fermi$ Large Area Telescope 3FHL catalog.
Out of 200 unidentified sources, we selected 110 sources for which archival $Swift$-XRT  observations were available, 52 of which were found to have exactly one X-ray counterpart within the 3FHL 95\% positional uncertainty. In this work, we report the X-ray, optical, IR, and radio properties of these 52 sources using positional associations with objects in various catalogs. The $Wide$-$field$ $Infrared$ $Survey$ $Explorer$ color-color plot for sources suggests that the most of these belong to the blazar class family. The redshift measurements for these objects range from $z=$0.277 to $z=$2.1. 
 Additionally, under the assumption that the majority of these sources are blazars, three machine-learning algorithms are employed  to classify the sample into flat spectrum radio quasars or BL Lacertae objects. These suggest that the majority of the previously unassociated sources are BL Lac objects, in agreement with the fact the BL Lac objects represent by far the most numerous population detected above 10\,GeV in  3FHL.

 \end{abstract}
\section{Introduction}
The $\gamma$-ray sky provides us with a unique opportunity to reveal the nature of the most extreme environments in the universe. The Energy Gamma Ray Experiment Telescope \citep[EGRET;][]{Fichtel1993} on the Compton Gamma Ray Observatory (CGRO) performed a survey of the $\gamma$-ray sky above 50\,MeV, revealing multiple high-energy astrophysical phenomena, such as active galactic nuclei (AGN), Supernova remnants, Gamma-ray bursts, pulsars. The large area telescope \citep[LAT;][]{Atwood2009} aboard the {\it Fermi} satellite was launched in 2008 and revolutionized this area of astrophysics by detecting thousands of sources in the $\gamma$-ray energy band. The latest released, broad-band, all-sky LAT catalog \citep[3FGL;][]{Acero2015} consists of 3033 sources detected in the energy range 0.1-300 GeV. Other two catalogs \citep[1FHL and 2FHL;][]{Ackermann2013a,Ackermann2016b} exploited the high-energy $\gamma$-ray sensitivity of {\it Fermi}-LAT, reporting sources detected above 10 GeV (514 objects) and 50 GeV (360). 
The most recently published 3FHL {\it Fermi}-LAT catalog \citep{Ajello2017} represents a significant upgrade of the 1FHL catalog and lists 1556 sources detected between 10\,GeV and 2\,TeV, utilizing 7 years of LAT data. 

While the  majority ($\sim$78\,\%) of the 3FHL sources has already been associated with a counterpart, 200 objects in this catalog lack any information on association or classification. The knowledge of the properties of these extremely high-energy sources is fundamental for the studies  of extragalactic background light \citep[EBL,][]{Dominguez2015} and to constrain the origin of the extragalactic $\gamma$-ray background \citep[EGB, see e.g.][]{Ajello2015,Ackermann2016a}. 
Furthermore, the 3FHL catalog will likely represent the main reference for future observations with the upcoming Cherenkov Telescope Array \citep[CTA;][]{Hassan2017}.\\
 The biggest challenge in finding associations to $\gamma$-ray sources are the rather large positional uncertainties ($\sim$ arcminutes). This issue can be minimized by conducting X-ray observations of these $\gamma$-ray source fields, which has been done in the past by various authors \citep[see][for recent studies with this approach]{Stroh2013,Parkinson2016,Paiano2017}. In the context of this work, we utilize observations performed with the X-ray Telescope (XRT) telescope mounted on the Neil Gehrels {\it Swift} Observatory \citep{Gehrels2004}. 
This paper is organized as follows: Section 2 describes the source selection criteria and analysis procedure for the $Swift$-XRT data, Section 3 lists the catalogs (radio/IR/optical) and the procedure used to find likely associations these 3FHL objects. Section 4 explains the machine learning methods employed to further classify these  sources. Section 5 describes the results of the classification process and Section 6 comprises of the discussion and conclusions based on our analysis. 
\section{$Swift$-XRT Data Selection}
\label{sec:obs}

Firstly, we queried the HEASARC\footnote{https://heasarc.nasa.gov/} database for {\it Swift} satellite observations of the fields corresponding to the unassociated 200 3FHL sources and we found 110 3FHL source fields observed with $Swift$-XRT. For 48 fields we found multiple observations. All these observations  were stacked before proceeding for the $Swift$-XRT analysis.  All the XRT data reduction processes, i.e., summing data from different observations, creating images and spectra for all the sources were performed with the online $Swift$-XRT product builder\footnote{http://www.swift.ac.uk/user\_objects/index.php} following the methods described in \citet{Evans2007, Evans2008}. This procedure was performed using standard tools in HEASOFT version 6.19. All the generated images were investigated to identify X-ray sources within the 95\% confidence interval for the {\it Fermi}-LAT positions. In three cases (3FHL J0316.5-2610, 3FHL J1248.8+5128, 3FHL J2042.7+1520) we found exactly one X-ray counterpart outside but close to the 95\% uncertainty (within 2 arcmin) and we have included these in our final sample. For another source, 3FHL J1553.8-2425, we found two bright counterparts, one within the 95\% region and one known cataclysmic variable about 2.5 arcmin outside the Fermi uncertainty region. We proceeded by including the counterpart inside the 95\% region as the associated source in our sample. In 55 fields observed with XRT, no source was detected within the 3FHL uncertainty radius. We point out that the 3FHL sources, where no X-ray counterpart was detected, have on average smaller exposure times (mean 1300 sec, median 1800 sec) than the ones where an X-ray source was detected (mean 5000 sec, median 4000 sec). Furthermore, images with more than one source ( three total) were excluded from the sample and left to a future study. These criteria lead to a total number of 52 X-ray candidate counterparts in our final sample. Each image was then investigated manually to estimate a rough position of the object, which was then provided in the above mentioned online tool to calculate the exact counterpart centroid employing the standard position method. These localized X-ray positions are listed in Table~\ref{tab:swift}. Spectral fitting was performed using  XSPEC version 12.9.1 \citep{Arnaud1996} utilizing the source and background files generated by the online tool. All the spectra were fitted with a power law in conjunction with the Tuebingen-Boulder ISM absorption model ({\tt\string tbabs}). The Galactic column densities were obtained with the HEASoft online tool\footnote{https://heasarc.gsfc.nasa.gov/cgi-bin/Tools/w3nh/w3nh.pl} \citep{Kalberla2005}.The resulting parameters from this analysis are presented in Table~\ref{tab:swift}.

\section{Correlation with other databases}\label{sec:data}
\begin{figure*}[]
 \includegraphics[trim=20.0cm 3.0cm 20.0cm 3.0cm,width=\columnwidth]{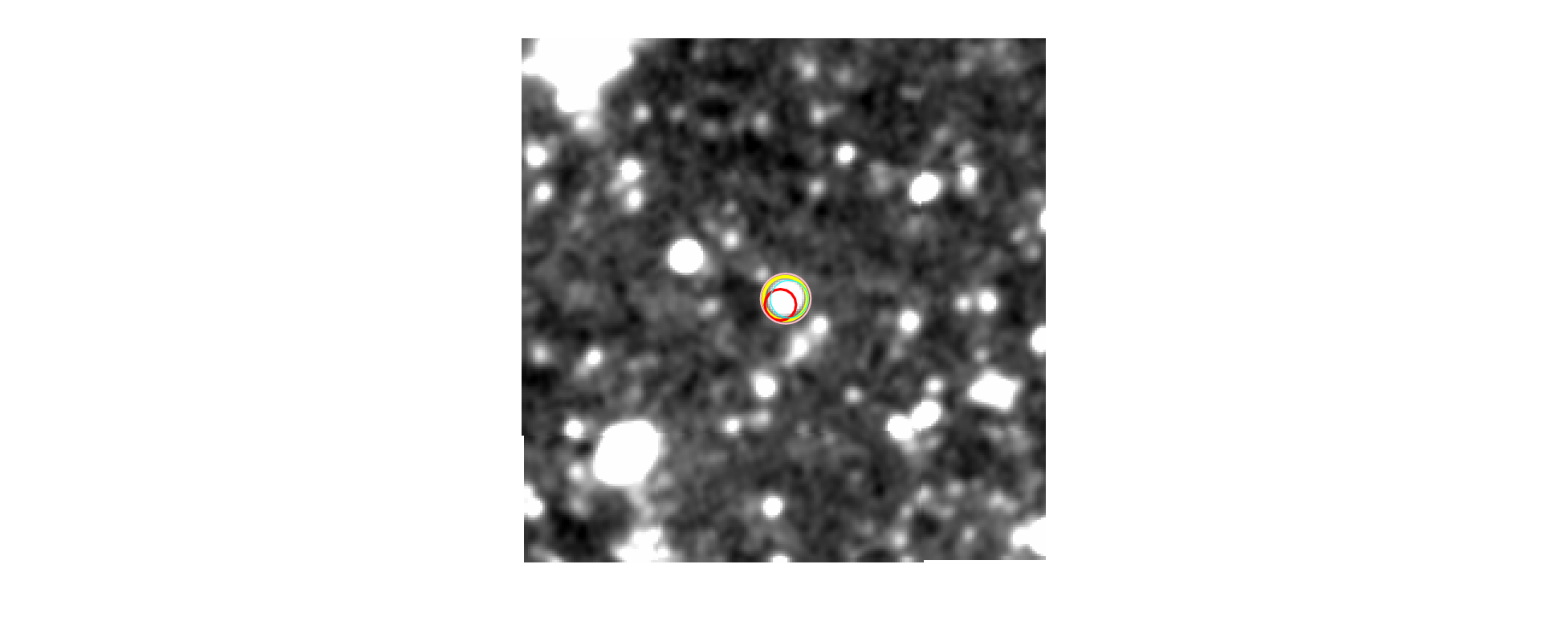}
 \caption{\footnotesize Optical image centered on 3FHL J0121.9-3917 position  ( 2.5 $\times$ 2.5 arcmin in size). The positions of counterparts from different catalogs are show with circles of different colors, as follows. NVSS: red; $Swift$-XRT: cyan; 2MASS: magenta; WISE: yellow. The size of each circle is adjusted to display all the catalogs clearly.}
 \label{fig:corr}
 \end{figure*}

We used the $Swift$-XRT positions to cross-correlate our sources with multiple catalogs, selected in different bands. See Fig.~\ref{fig:corr} Based on the $Swift$-XRT typical positional accuracy, we allowed a maximum position uncertainty of 5\asec, except for the cross-correlation with the ROSAT catalog \citep[1RXS,][see below]{Voges1999}. The resulting associations are presented in Table~\ref{tab:corr}. The different catalogs used for the association are reported in the following sections.
\subsection{1RXS}
The ROSAT All-Sky Survey Source Catalogs for bright and faint sources \citep[1RXS][]{Voges1999,Voges2000} contain 18806 and 105924 sources, respectively. The positional uncertainties for these sources are of the order of 30\,\farcs, therefore, we compare the 1RXS positions with the $Swift$-XRT positions allowing a maximum uncertainty of 30\,\farcs to find the possible associations. This lead to 8 positional correlations, as listed in Table~\ref{tab:corr}. The 1RXS positional error is reported in parenthesis for each association in this table. 

\subsection{BZCat} The 5th Roma-BZCat catalog is the largest known blazar sample \citep{Massaro2015a,Massaro2016}. Three out of 52 sources were found spatially coincident with BZCat sources within 5\asec and all of them are classified as BL Lacs: 3FHL J0316.5-2610, 3FHL J1248.8-5128 and 3FHL J1553.8-2425 with redshifts, $z$= 0.443, 0.351 and 0.332, respectively.
\subsection{AllWISE}

We cross-correlated our sample with AllWISE, the complete Wide-field Infrared Survey Explorer (WISE) point source catalog \citep{Cutri2013}, and WIBRaLS, a catalog of radio-loud candidate $\gamma$-ray emitting blazars with WISE colors similar to the colors of confirmed $\gamma$-ray blazars \citep{DAbrusco2014}. 39 sources were found coincident with AllWISE positions and 3 identified as BL Lacs in the WIBRals catalog, which are the same  found in BZCat. (see Table~\ref{tab:corr}).  31 of these sources were detected in all the W1, W2, W3 and W4 filters (3.4, 4.6, 12 and 22 $\mu$m, respectively). However, only upper limits for W3 and W4 were provided for 8 sources. See Fig.~\ref{fig:wise}
\subsubsection{WISE color index classification} 

\citet{Massaro2012} introduced a method to identify blazars of uncertain type using a four-filters WISE color-color diagram \citep{Wright2010}. These authors identified a particular region in the diagram, which separates blazars from other source classes. They termed this region the WISE blazar strip (WBS) \citep[see Fig~1, 2 in][]{Massaro2012}. Moreover, they found that BL Lacs 
and FSRQs
follow a bimodal distribution, such that the former occupy the bluer part of the color-color diagram. Utilizing the information obtained from the spatial correlation with the AllWISE catalog in our sample of 39 sources with WISE counterpart, we compare their position on the WISE color-color plot with that of the 915 known blazars from the 3FHL catalog. 
It is evident from Fig.~\ref{fig:wise} that $>$80\% of our sources lie within the WISE Gamma-ray Strip Projections \citep{Massaro2012} for BL Lacs and FSRQs; and the majority of these occupy the BL Lac region.

\subsection{Million Quasar catalog}
The catalog published by \citet{Flesch2017} presents type I and II QSOs , AGNs and BL Lacs reported in various catalogs from the literature before 21 June, 2016. This list also includes the candidates based on the SDSS photometric quasar catalogs. Ten sources in our sample have an association in the Million Quasar catalog: two were coincident with QSO type I, three with BL Lacs and the other five with possible QSOs with likelihood $>$85\%. Finally, the match to the Million Quasar catalog yielded redshifts for two sources (see Table~\ref{tab:corr}). These redshifts were derived using the photometric method utilizing the SDSS DR12 catalog \citep{Alam2015}, details of which are provided in the Half Million Quasar  catalog \citep{Flesch2015}. Seven of these ten sources are also spatially coincident with a radio counterpart. 
\subsection{6dFGS}
The 6dF Galaxy Survey (6dFGS) catalog \citep{Jones2009} provides the redshift map of the Southern hemisphere for nearby objects ($z$ $\lesssim$ 0.1). None of the sources in this catalog are positionally coincident with our sample sources within a 5$^{\prime\prime}$ uncertainty radius circle.
\subsection{Radio catalogs}
\subsubsection{NVSS}
The NRAO VLA Sky Survey (NVSS) \citep{Condon1998} constitutes the radio observations of celestial objects with declination greater than -40 $\deg$ at 1.4 GHz. 
The radio positions of the 22 counterparts of the 3FHL sources is presented in Table~\ref{tab:corr}. 
\subsubsection{SUMSS}
The Sydney University Molonglo Sky Survey  \citep[SUMSS;]{Mauch2003} with radio observations at 843 MHz consists of Southern Hemisphere objects. Two sources in our sample have a radio counterpart in this catalog (see Table~\ref{tab:corr}).
\subsubsection{2WHSP}
 \citet{Chang2016} assembled the largest known catalog of WISE High Synchrotron Peak blazars (2WHSP) which comprises 1691 sources. These authors cross-correlated AllWISE catalog with various other wavelength surveys (radio, IR, X-ray). Utilizing this multiwavelength information, they identified blazars, calculated their peak synchrotron frequencies ($\nu_{syn}^{pk}$) and listed the HSPs ($\nu_{syn}^{pk}$ $>$ 15 Hz) in the 2WHSP catalog. We found 9 sources from our sample to be spatially coincident with HSP blazars in this catalog. The 2WHSP identifications of these sources are provided in Table~\ref{tab:corr}.
\subsection{Miscellaneous}
Finally, we checked for potential counterparts using the two largest online databases of astronomical objects, i.e., SIMBAD \citep{Wenger2000} and the NASA/IPAC Extragalactic Database (NED)\footnote{https://ned.ipac.caltech.edu/}. We found the following possible associations, which were not found in any of the above mentioned catalogs:

\textbf{3FHL J0438.0-7328, 3FHL J1249.2-2809}: These two sources are positionally coincident with galaxies LEDA 255538 and LEDA 745327, respectively. These associations are derived from HYPERLEDA, a catalog of about one million galaxies brighter than B=18\,mag \citep{Paturel2003}. No redshift information was found for these two objects. 

\textbf{3FHL J1234.8-0435}: This 3FHL source spatially coincides with a galaxy listed in the 2dF Galaxy Redshift Survey \citet{Colless2001}.
The redshift, $z$=0.277, provided by this catalog was obtained from both absorption and emission features.

\section{Classification with machine learning}

Multiwavelength data analysis is typically required for every {\it Fermi} detected source to be correctly classified. This process is highly time consuming and the lack of this information has led to an increasing fraction of unidentified sources in every new {\it Fermi} catalog release.
However, $\sim$ 80\% of the objects in the 3FHL catalog are associated with blazars (FSRQ, BL Lac or BCU), and this fraction increases to $\sim$ 90\% if  sources along the Galactic plane ($|b|\leq 10^{\circ}$) are not considered. In our sample, 36 out of 52 sources are high-latitude objects, therefore we assume that all these 36 unknown sources in our sample are blazars. This assumption is also justified by the fact that, as seen in Fig.~\ref{fig:wise}, most of the sources are coincident with the WISE-blazar strip, except three outliers: 3FHL J0427.5-6705 (W3 measurement S/N=0.1, no radio), 3FHL J1650.9+0430 (W3 measurement S/N =0.1, no radio) and 3FHL J1958.1+2437.\\
Although it is quite evident from Fig.~\ref{fig:wise} that most of our sources lie within the blazar region, in particular, BL Lac region, these results are based on only two parameters. We thus further analyze our findings by employing various other properties, to refine the way in which we differentiate BL Lacs from FSRQs.
In order to accomplish this multi-parameter space classification, we employ three different machine learning algorithms. The parameters employed in these methods were derived from the 3FGL, 1SXPS \citep{Evans2014} and AllWISE catalogs, and are discussed in detail in Section~\ref{sec:params}.
Various machine learning techniques have been successfully applied to $Fermi$ unidentified sources, e.g., \citet{Ackermann2012,Mirabal2012,Mirabal2016,Parkinson2016,Salvetti2017}. From a wide variety of available methods used by these  authors, we chose to apply three most commonly employed methods: Decision Tree \citep{Quinlan1990}, Support Vector Machines \citep{Hearst1998} and Random Forest \citep{Breiman2001}. 
 \subsection{Decision Tree}
 The decision tree classifier (DT) is an example of supervised machine learning algorithm which separates a dataset into two or more categories based on certain parameters associated with the input data. The data is continuously split into nodes and branches until every data point is assigned to one or the other category. The decision of splitting into separate nodes is based on the Gini index, an impurity measurement. The Gini impurity parameter provides a measurement of the probability of incorrectly labeling a randomly chosen element in the given dataset. The decision tree algorithm works towards minimizing this value and splits the sample into branches until this index reaches zero. Mathematically, it is defined as 
 $$G=1-\Sigma_{i=1}^{J}p_{i}^{2}$$
 which calculates the Gini impurity for a dataset with $J$ categories with $p_{i}$ being the fraction of items labeled with category $i$ in the sample. Higher values of G imply higher inequality between two classes for a given parameter. The decision tree is split until the Gini index reaches a minimum value equal to zero, thereby assigning  a particular class to the underlying items. This method employs a dataset with known classification as training dataset and trains the classifier. The accuracy obtained for a trained dataset is calculated to evaluate its usage on a sample with no classification. 
 
 \subsection{Support Vector Machines}
 Support Vector Machine (SVM) is another supervised learning method for separating a dataset into two categories. The underlying principle for this method is that for any data point, $i$ or $j$ (two categories), one or a set of maximum margin hyperplane are found such that the distance between this plan and the nearest point in either category is maximized. 
 
Mathematically, a hyperplane for a set of points with category , $i$ (say $\vec{x}$) is defined as following:
 $$\vec{w}.\vec{x}-b=0,$$
 such that the parameter $\frac{b}{||\vec{w}||}$ defines the distance of this plane from the origin along $\vec{w}$, where $\vec{w}$ is the normal vector to the hyperplane. This is an example of linear kernel classification for SVM. A non-linear SVM employs polynomial or rbf kernels to classify any dataset for higher dimensions. 
In the context of our work, we employ a polynomial kernel and a non-linear SVM to classify the sample into two kind of blazars, i.e., BL Lacs or FSRQs. 
\subsection{Random Forest}
 A random forest is one of the most commonly employed supervised machine learning method used for both classification and regression analysis. A random forest classifier operates as an ensemble algorithm based on the principle of a decision tree  classifier. This method constructs various decision tree algorithms and assigns a class to a source for every iteration. An aggregate of these predicted classes is assigned as the final resulting class for that particular source. This method has an advantage over running a single decision tree, since it utilizes the multitude of decision trees, thereby solving the problem of overfitting \citep{Hastie2009}, which is usually observed in the latter case. We employ this method to classify our sample into BL Lacs and/or FSRQs. This yields probabilities for each source to be associated as a BL Lac or an FSRQ.
\subsection{Sample and Parameter Section}\label{sec:params}

We employed the DT classifier, SVM classifier and Random Forest implemented in {\tt\string sklearn0.20.0} library \citep{Pedregosa2011}  in {\tt\string python 2.7} on a  sample of 152 3FHL blazars (115 BLLacs and 37 FSRQs). This sample was chosen as a subset of all the BL Lacs and FSRQs in the 3FHL catalog for which all the six parameters listed in  Table~\ref{tab:pars} were available. The reason for the selection of these six properties was based on the fact that these have been observed to distinguish BL Lacs from FSRQs. In general, BL Lacs exhibit harder spectrum in Gamma-rays (e.g. \citet{Abdo2010b,Ackermann2015b}) and softer in X-rays (e.g. \citet{Donato2001}) as compared to FSRQs , therefore, we select the spectral indices in Gamma-ray (3FHL and 3FGL) and X-ray. The WISE colors, as already discussed in the text, clearly differentiate the two classes of blazars.
FSRQs can be distinguished from BL Lacs on the basis of variability. In the 3FHL catalog \citep{Ajello2017}, a parameter called Variability Bayes Blocks is provided, which lists the number of Bayesian blocks from variability analysis. The values of this parameter range from -1 to 15, where -1 implies no variability and 15 implies high variability. FSRQs exhibit higher values for this parameter, implying higher variability as compared to BL Lacs. 
This sample was divided into training and test datasets in order to check the accuracy of the method employed. The training and test datasets comprised of 102 blazars (77 BL Lacs and 25 FSRQs) and 50 blazars (38 BL Lacs and 12 FSRQs), respectively. Since the total sample contains $\sim$ 75\% BL Lacs and only 25\% FSRQs, which being highly imbalanced could yield inaccurate results biased towards the major class, when a machine learning method is applied. We, therefore, employed a technique called SMOTE (Synthetic Minority Over-sampling Technique) \citep{Chawla2002}. This method creates synthetic minority class using k nearest neighbors algorithm, thereby generating equal number of sources in each class. An an example, in this case, the training dataset has 77 BL Lacs and 25 FSRQs. Implementation of SMOTE method generated 52 synthetic FSRQs, thereby balancing the two classes (77 sources for each class) before application of a classification method.
\section{Results}
\label{sec:results}

 The resulting decision tree from the training sample is shown in Fig.~\ref{fig:dttrainer}. We employed this trained classifier on the test dataset (38 BL Lacs and 12 FSRQs) which yielded an accuracy of 86\%. This classifier was then applied to the unknown 3FHL sample, which yielded results suggesting that 31 out of the 36 high-latitude unassociated sources are BL Lacs and the rest are FSRQs. 
For SVM, a receiver operating characteristic curve (ROC) is used to evaluate the accuracy of this binary classifier and it is shown in Fig.~\ref{fig:svm_roc}. The ROC is constructed by plotting the true positive rate (TPR, number of correct positive results) against the false positive rate (FPR, number of incorrect positive results) at various thresholds. The classification accuracy is $\sim$ 90\% which was evaluated as the area below the curve for the given sample. The SVM analysis on our sample suggests that all the unknown sources are likely BL Lacs. 
The Random Forest classifier yielded results consistent with the SVM classifier suggesting that all the unassociated 36 high-latitude sources are BL Lacs. The accuracy obtained on the test sample in this case was 98\%. The receiver operating characteristic curves for both the training and the test sample are shown in Fig.~\ref{fig:rf_roc}. 
A comparison of the results yielded by all these methods is displayed in Table~\ref{tab:mlcomp}.

\section{Discussion and Conclusions}

\begin{figure*}[]
  	\includegraphics[trim=2.0cm 0.0cm 0.0cm 0.0cm,angle=0,width=\columnwidth]{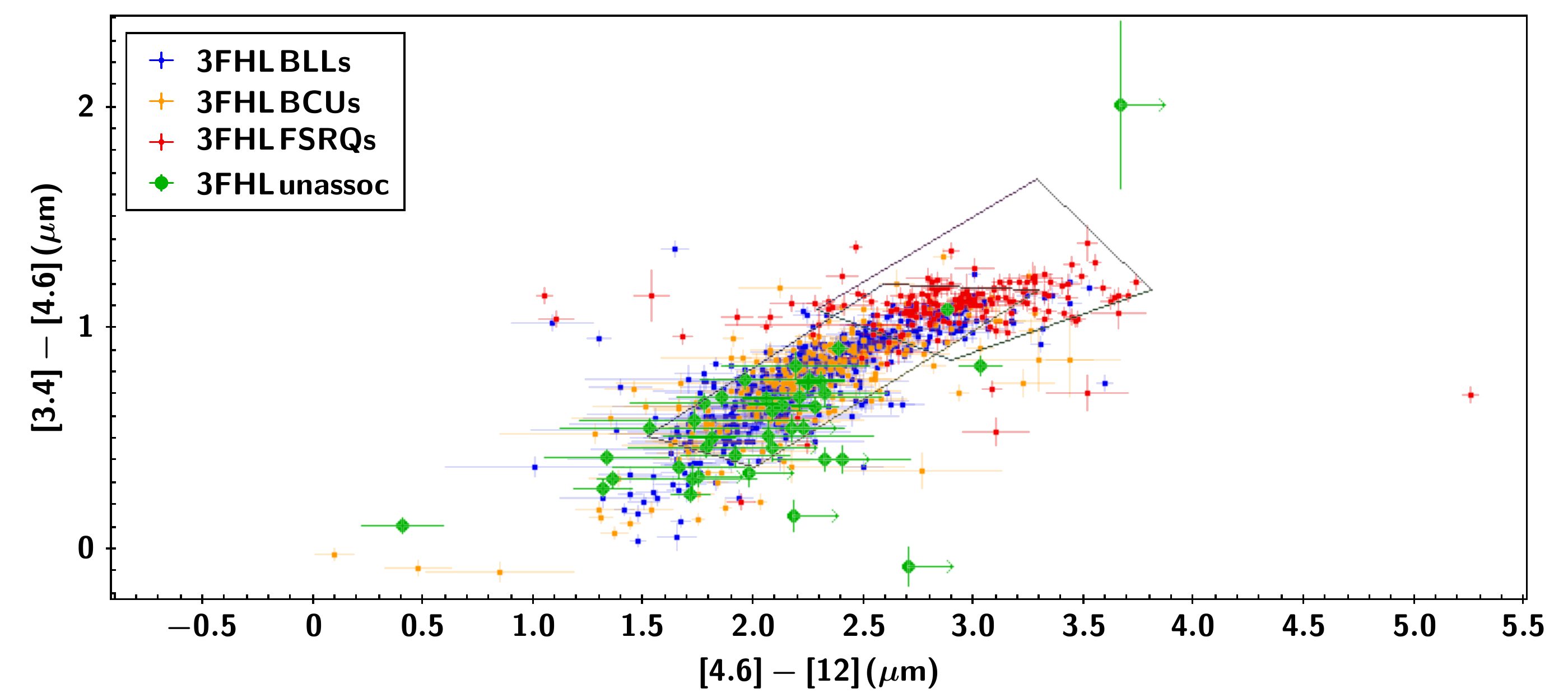}
    \caption{\footnotesize WISE color-color diagram for all the known blazars in the 3FHL catalog following the technique introduced by \citet{Massaro2012}. The $blue$, $red$, $orange$ circles represent BL Lacs, FSRQs and BCUs, respectively. The $green$ circles represent our sample of 52 sources from the 3FHL data. The two boxes represent the WISE Gamma-ray Strip Projection for BZBs (BL Lacs) ($left$) and BZQs (FSRQs) ($right$), as described in \citet{Massaro2012}. This plot clearly suggests that the majority of the sources in our sample are likely to be BL Lacs. The $green$ arrows represent 3FHL unassociated sources for which only an upper limit was provided for the W3 (12 $\mu$m) filter. }
 
  	\label{fig:wise}
  \end{figure*}

\begin{figure*}[]
 	\includegraphics[trim=2.0cm 1.0cm 0.0cm 0.0cm,angle=0,width=\columnwidth]{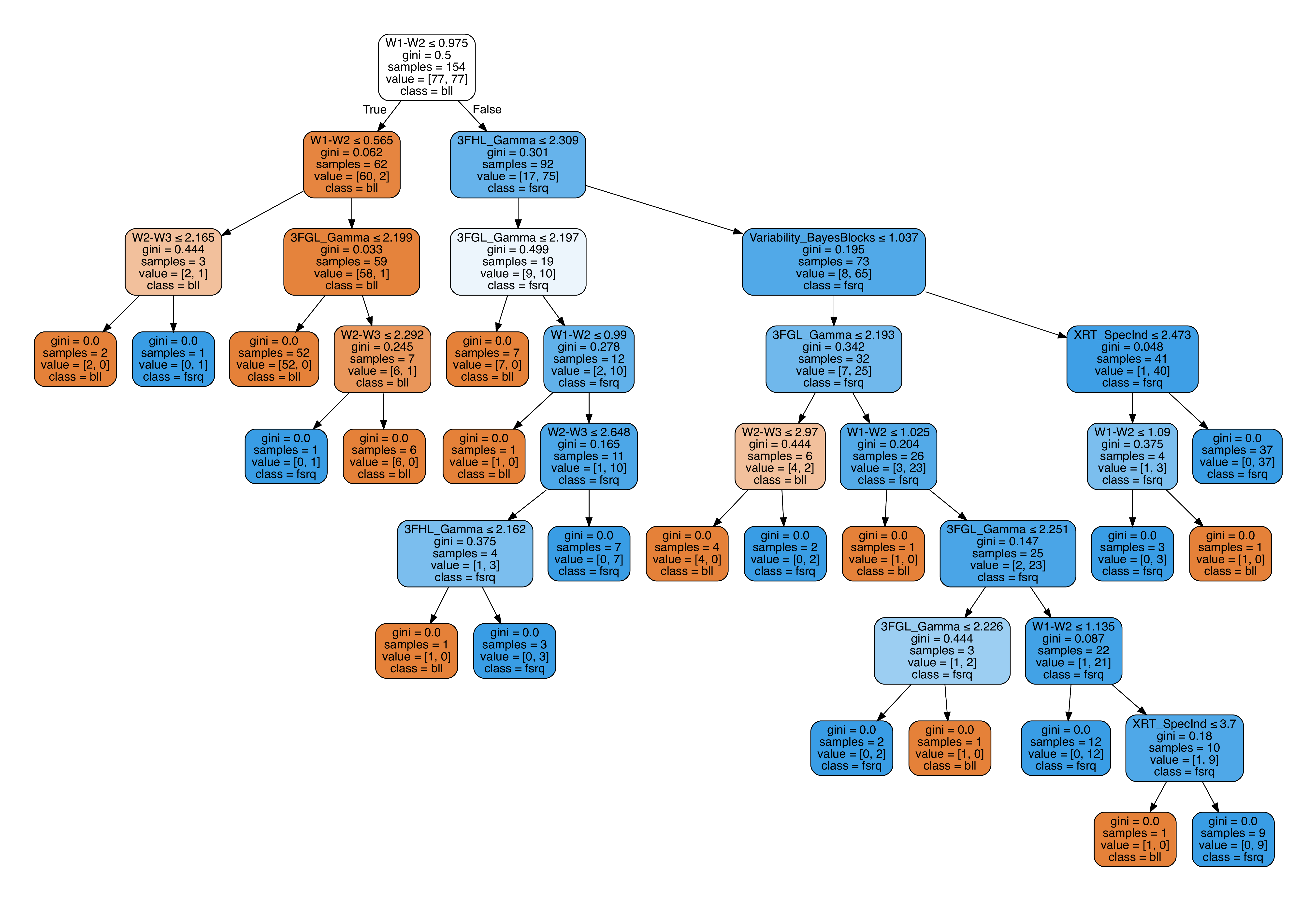}
 	\caption{\footnotesize  The Decision Tree  for the training data set which contains 77 BL Lacs and FSRQs, each, employed six parameters based on known distinct properties of two kind of blazars.}
 	\label{fig:dttrainer}
 \end{figure*}

\begin{figure*}[]
\centering 
\subfloat[]{%
\hspace{-4.5cm}
  \includegraphics[width=0.5\textwidth,height=6.5cm,trim=12mm 10mm 25mm 3.8mm, clip=true]{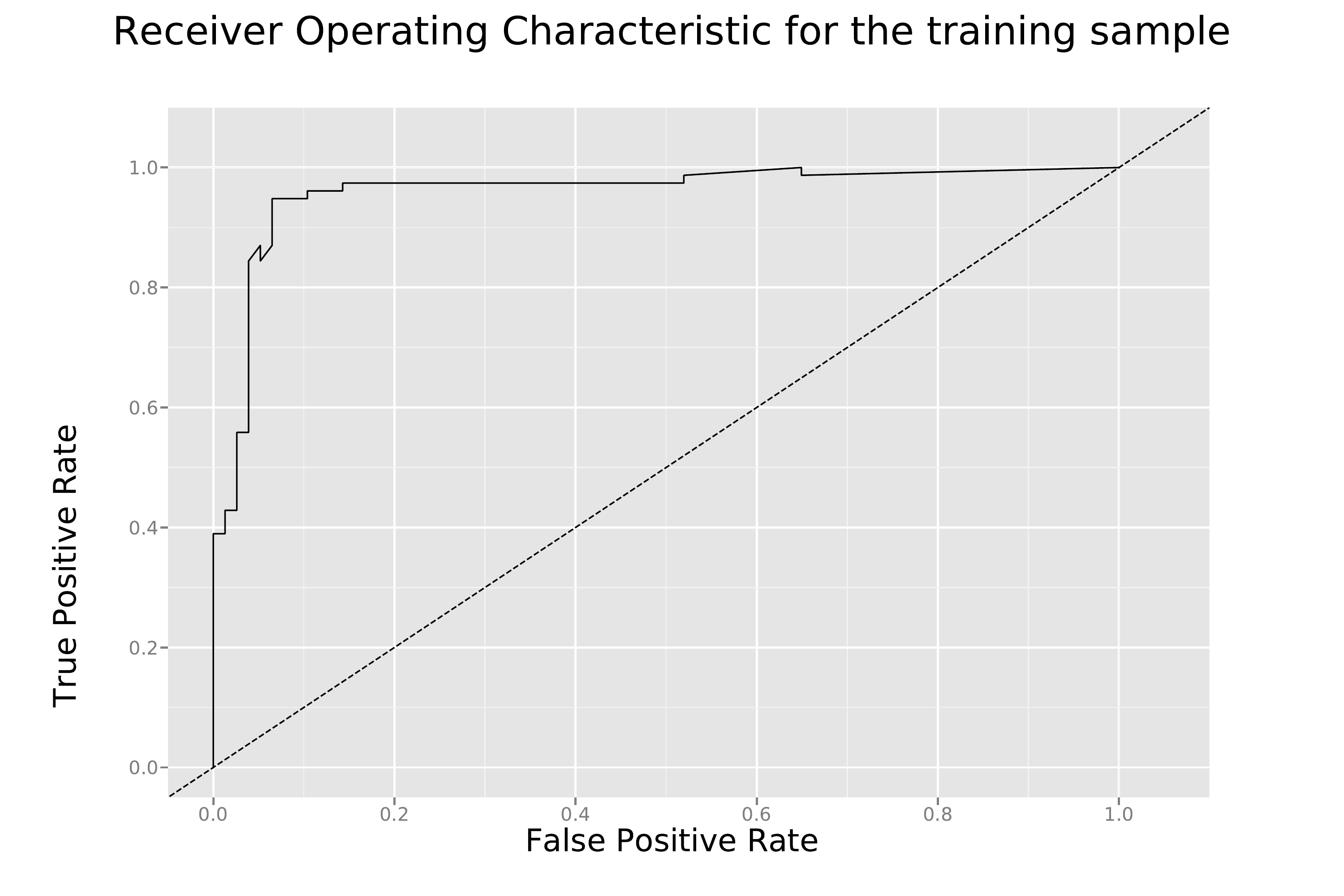}%
}\qquad
\hspace{-0.8cm}
\subfloat[]{%
  \includegraphics[width=0.5\textwidth,height=6.5cm,trim=10mm 10mm 25mm 3.8mm, clip=true]{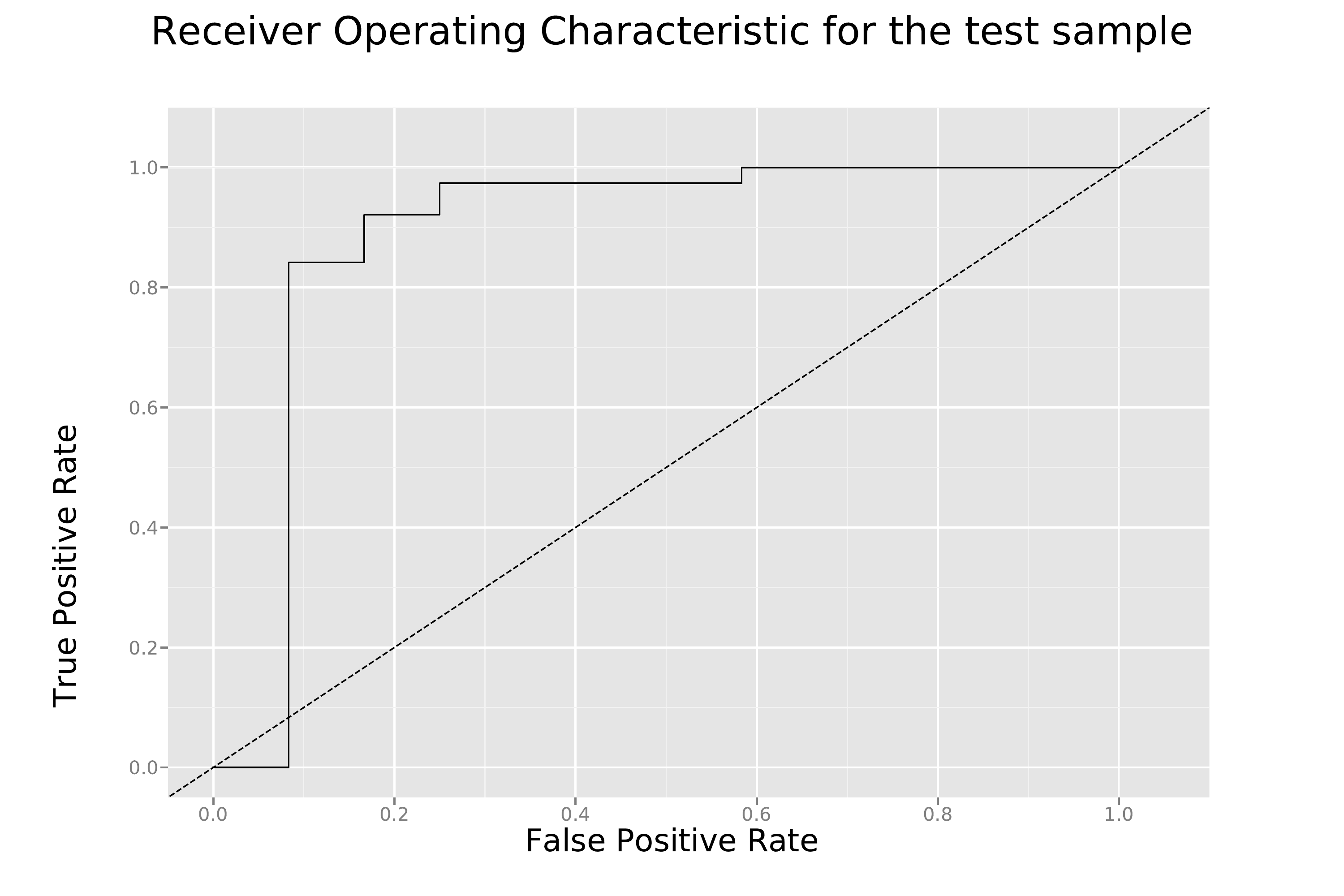}%
  \hspace{-4.9cm}
}
\caption{\scriptsize  The SVM method generated ROC curves for the (a) training sample and the (b) test sample. The test sample yielded an accuracy score 90\%. The straight diagonal line in both the plots represent the non-discriminatory curve, i.e., if the data points lie on/below this line, the analysis would yield non-diagnostic results.}
\label{fig:svm_roc}
\end{figure*}

\begin{figure*}[]
\centering 
\vspace{-2.0cm}
\subfloat[]{%
\hspace{-5.5cm}
  \includegraphics[width=0.52\textwidth,height=6.5cm,trim=12mm 15mm 26mm 3.8mm, clip=true]{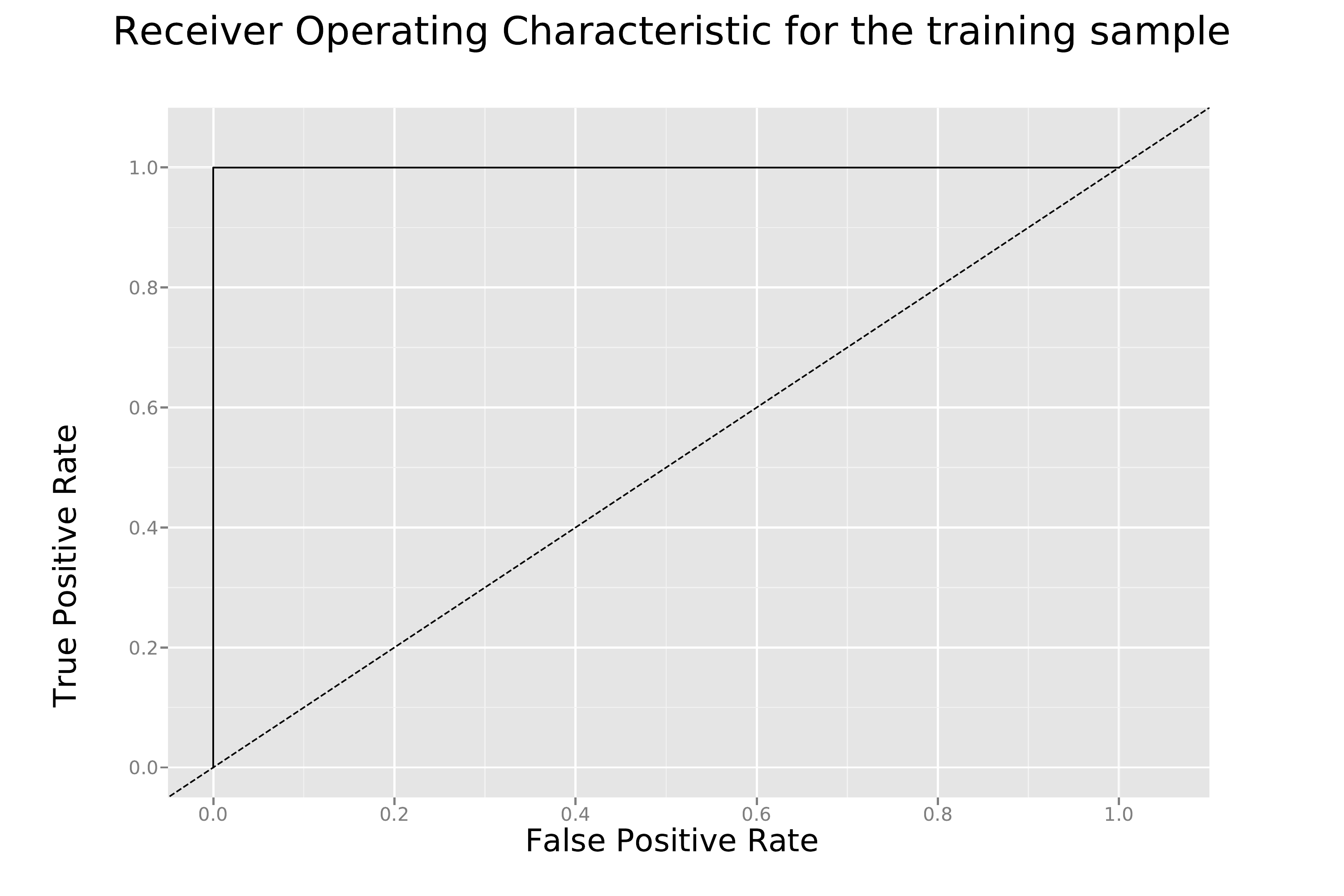}%
}\qquad
\hspace{-0.8cm}
\subfloat[]{%
  \includegraphics[width=0.52\textwidth,height=6.5cm,trim=10mm 15mm 35mm 3.8mm, clip=true]{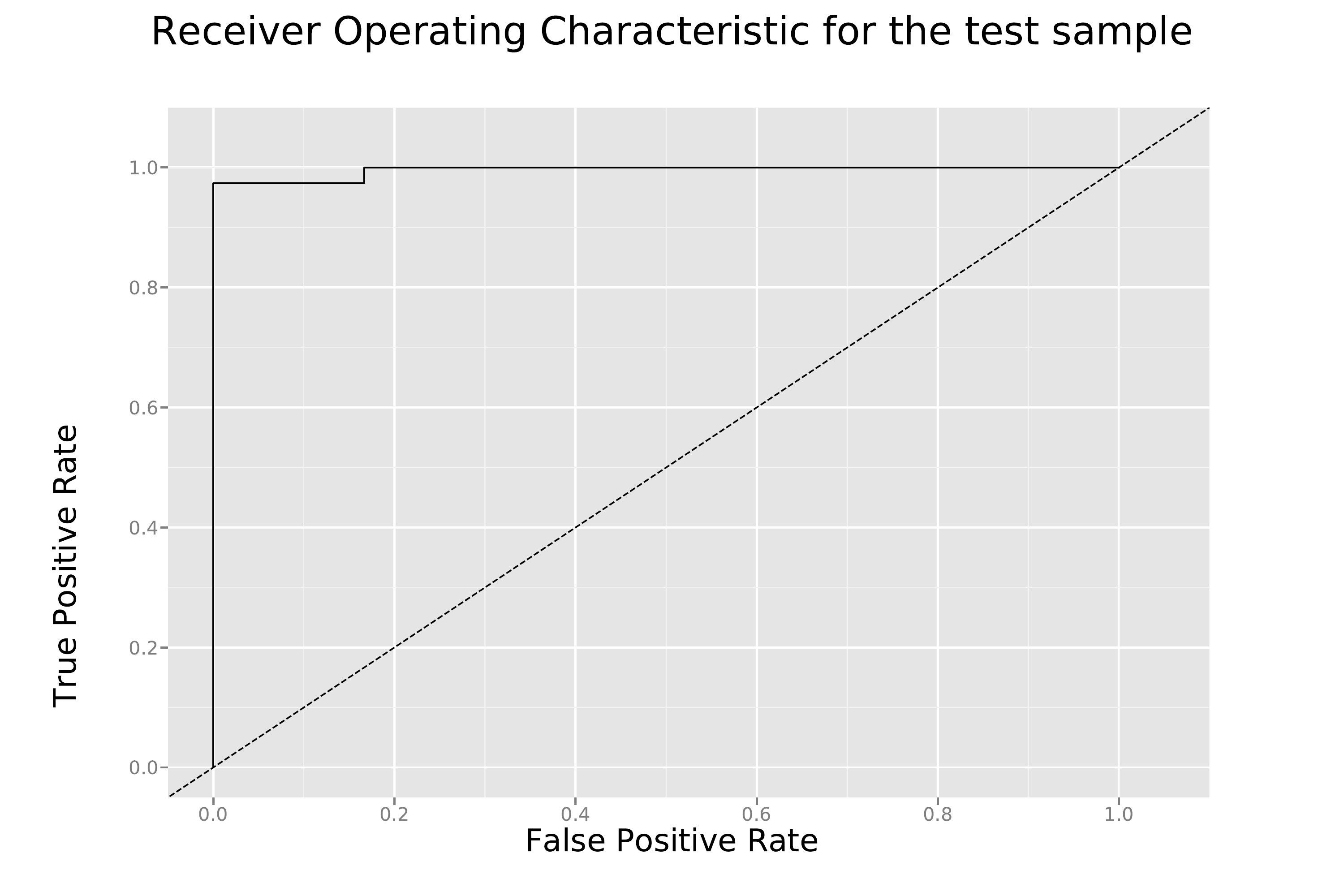}%
    \hspace{-5.5cm}
}
\caption{\scriptsize The ROC curves from the Random Forest method for the (a) training sample and the (b) test sample. The test sample yielded an accuracy score 98\%. The straight diagonal line in both the plots represent the non-discriminatory curve, i.e., if the data points lie on/below this line, the analysis would yield non-diagnostic results.}
\label{fig:rf_roc}
\end{figure*}

The immediate objective of this work was to identify the nature of unassociated sources reported in the latest $Fermi$ high energy catalog, the 3FHL. In an attempt to find associations to these sources, $Swift$ HEASARC archive was used to derive the accurate positions of the sources for which data were available, which lead to a sample of 52 3FHL unassociated sources with a single bright X-ray counterpart. The X-ray source positions were cross-matched with various catalogs from radio to X-ray wavelengths (see Section \ref{sec:data}), leading to the identification of the likely counterpart for 12 out of the 52 objects (6 out of 52 objects are identified as QSOs, 3 as BL Lacs and 3 as galaxies).
Six of these 12 sources also have confirmed redshift measurements ranging from $z$=0.277 to $z$=2.1. In addition, the WISE color-color plot, as shown in Fig.~\ref{fig:wise}, suggests that majority of the 3FHL sources are likely blazars. Moreover, 90\% of the high-latitude $|b|\geq$ 10$^{\circ}$ objects in the 3FHL sample are associated with the blazar population. 36 objects from our source sample are high-latitude objects and assuming that this subsample comprises only blazars, we employed machine learning techniques (DT, SVM and RF) to classify these objects into two kind of blazars, i.e., BL Lacs and FSRQs. The DT classifier yielded results showing that 31 of the high-latitude  sources are BL Lacs. The SVM and the RF classifier predict that  all these sources are BL Lacs. For details, please see Table~\ref{tab:mlcomp}. The inconsistency between the results from DT vs SVM/RF could be attributed to potential overfitting in the former method as discussed earlier. \\
In nutshell, this work provides classification for 36/200 sources, which reduces the incompleteness of the 3FHL by 18\%. While the redshift info is scarce (12\%), our group is working on an optical spectroscopic campaign to observe these unassociated sources with 4m and 8m class telescopes, to obtain redshifts for a significant fraction of them and confirm their nature (see \citet{Marchesi2018}, where the first results of this campaign are reported). In addition, our recent successful proposal (Swift Cycle 14, prop ID 1417063 PI: Ajello) in an effort to obtain more XRT sources for the unknown/unassociated 3FHL sources is currently in progress. All these continuing studies will drastically reduce the incompleteness of the 3FHL catalog in a few years timescale. \\

\onecolumn


 \begin{deluxetable}{ccccccccccc}

 	\tabletypesize{\scriptsize}
    \rotate
 	\tablecolumns{11}
 	\tablewidth{0pt}
 	\setlength{\tabcolsep}{0.04in} 
 	\tablecaption{\large $Swift$-XRT Analysis \label{tab:swift}}
 	\tablehead{
 		\colhead{3FHL } &
 		\colhead{RA2000$^a$} &
 		\colhead{Dec2000$^b$} &
 		\colhead{Xray RA$^c$} &
 		\colhead{Xray Dec$^d$} &
         \colhead{Exposure time} &
        \colhead{Count Rate} &
 		\colhead{$\Gamma_{X}^e$} &
 		\colhead{N$_{H}^f$} &
 		\colhead{Flux$^g$} &
 		\colhead{$\chi^2/d.o.f.$} \\
 		\colhead{ } &
 		\colhead{(hh:mm:ss)} &
 		\colhead{($^{\circ}:^{\prime}:^{\prime\prime}$)} &
 		\colhead{(hh:mm:ss)} &
 		\colhead{($^{\circ}:^{\prime}:^{\prime\prime}$)} &
 		\colhead{(sec)} &
        	\colhead{(cts s$^-1$)} &
            	\colhead{} &
 		\colhead{(cm$^{-2}$)} &
 		\colhead{(erg cm$^{-2}$s$^{-1}$)} &
 		\colhead{} }
 	\startdata
     	J0049.0+4224 & 00:49:05 & +42:24:12 & 00:48:59.14 & +42:23:47.40  & 4106& 1.216& 2.54$\pm$0.24 & 0.12 & 0.68 & 48.31/47$^{*}$\\
 	J0121.9$-$3917 & 01:21:56 & $-$39:17:13 & 01:21:52.51 & $-$39:15:44.64  & 5736 & 0.080 & 1.79$\pm$0.08 & 0.02 & 3.45 & 19.21/19\\
 	J0156.2$-$2419 & 01:56:16 & $-$24:19:30 & 01:56:24.31 & $-$24:20:06.72  & 17280& 0.012 & 2.59$\pm$0.13 & 0.01 & 0.98 & 5.13/9\\
 	J0213.9$-$6950 & 02:13:58 & $-$69:50:12 & 02:13:58.44 & $-$69:51:38.16  & 4098 & 0.100 & 1.87$\pm$0.08 & 0.05 & 4.44 & 17.31/17\\
 	J0251.2$-$1830 & 02:51:13 & $-$18:30:30 & 02:51:11.31 & $-$18:31:15.57  & 7572 & 0.014& 2.03$\pm$0.16 & 0.03 & 0.64 & 68.78/88$^{*}$\\
 	J0316.5$-$2610 & 03:16:32 & $-$26:10:03 & 03:16:14.95 & $-$26:07:57.36  & 4885 & 0.152  & 2.52$\pm$0.06 & 0.01 & 4.82 & 18.90/30\\
 	J0350.4$-$5143 & 03:50:26 & $-$51:43:44 & 03:50:28.37 & $-$51:44:54.96  & 1708 & 0.160 & 2.08$\pm$0.13 & 0.01 & 6.26 & 10.12/10\\
 	J0350.8$-$2814 & 03:50:50 & $-$28:14:02 & 03:50:51.34 & $-$28:16:33.96  & 11010& 0.068 & 1.92$\pm$0.06 & 0.01 & 3.03 & 34.24/32\\
 	J0401.0$-$5355 & 04:01:03 & $-$53:55:11 & 04:01:11.45 & $-$53:54:56.52  & 904  & 0.020 & 1.95$\pm$0.33 & 0.01 & 0.78 & 11.55/15$^{*}$\\
 	J0427.5$-$6705 & 04:27:35 & $-$67:05:49 & 04:27:49.51 & $-$67:04:34.68  & 14500& 0.008 & \nodata & 0.04 & \nodata  & \nodata $\dagger$\\
 	J0438.0$-$7328 & 04:38:05 & $-$73:28:26 & 04:38:37.30 & $-$73:29:22.20  & 3868 & 0.007 & 2.10$\pm$0.42 & 0.09 & 0.19  & 13.89/25 \\
 	J0506.9+0323 & 05:06:55 & +03:23:42 & 05:06:50.09 & +03:23:59.28  & 17280& 0.012 & 2.47$\pm$0.15 & 0.07 & 0.57 & 3.7/7$^{*}$\\
 	J0541.1$-$4855 & 05:41:10 & $-$48:55:40 & 05:41:07.03 & $-$48:54:10.45  & 1224 & 0.008 & 2.04$\pm$0.76 & 0.03 & 0.47 & 3.8/8$^{*}$\\
 	J0559.6+3045 & 05:59:40 & +30:45:44 & 05:59:40.42 & +30:42:32.75  & 4680 & 0.004& 2.04$\pm$0.76 & 0.36 & 0.48 & 9.6/18$^{*}$\\
 	J0706.1+0247 & 07:06:08 & +02:47:55 & 07:06:10.86 & +02:44:50.82  & 2432 & 0.038 & 1.88$\pm$0.23 & 0.36 & 2.91 & 68.18/74$^{*}$  \\
 	J0725.7$-$0548 & 07:25:44 & $-$05:48:53 & 07:25:47.69 & $-$05:48:27.00  & 4872 & 0.036 & 2.37$\pm$0.18 & 0.26 &  2.39 & 10.22/6 \\
 	J0739.7$-$6720 & 07:39:45 & $-$67:20:09 & 07:39:27.74 & $-$67:21:37.04  & 3192 & 0.041 &  2.05$\pm$0.16& 0.09 & 2.36 & 82.00/96$^{*}$\\
 	J0747.7$-$4927 & 07:47:44 & $-$49:28:00 & 07:47:25.30 & $-$49:26:32.64  & 4086 & 0.023 & 2.44$\pm$0.18 & 0.09 & 1.18 & 66.75/71$^{*}$\\
 	J0813.7$-$0353 & 08:13:46 & $-$03:53:57 & 08:13:38.02 & $-$03:57:17.64  & 3157 & 0.066 & 1.71$\pm$0.15 & 0.05 & 3.37 & 5.89/7\\
 	J0820.2$-$2803 & 08:20:15 & $-$28:03:04 & 08:20:14.69 & $-$28:03:03.60  & 1763 & 0.022 & 2.49$\pm$0.35 & 0.27 & 1.95 & 26.96/36$^{*}$\\
 	J0928.5$-$5256 & 09:28:33 & $-$52:56:05 & 09:28:18.98 & $-$52:57:05.40  & 3579 & 0.006 & \nodata & 0.94 & \nodata & \nodata $\dagger$\\
 	J0937.8$-$1434 & 09:37:52 & $-$14:34:16 & 09:37:54.86 & $-$14:33:43.56  & 4480 & \nodata & \nodata & 0.05 &  \nodata & \nodata $\dagger$ \\
 	J1016.2$-$4245 & 10:16:14 & $-$42:45:34 & 10:16:21.02 & $-$42:47:24.36  & 3883 & 0.020& 2.39$\pm$0.19 & 0.05 & 0.89 & 54.36/65$^{*}$\\
 	J1024.5$-$4543 & 10:24:33 & $-$45:43:47 & 10:24:32.78 & $-$45:44:27.96  & 4158 & 0.057& 2.48$\pm$0.13 & 0.12 & 2.87 & 81.9/70\\
 	J1033.4$-$5033 & 10:33:26 & $-$50:33:39 & 10:33:32.28 & $-$50:35:28.68  & 3836 & 0.029& 2.25$\pm$0.17 & 0.19 & 1.78 & 83.36/98$^{*}$\\
 	J1034.8$-$4645 & 10:34:53 & $-$46:45:09 & 10:34:38.45 & $-$46:44:01.92  & 1733 & 0.063 & 1.97$\pm$0.19 & 0.14 & 3.40 & 3.36/4\\
 	J1047.9$-$3738 & 10:47:56 & $-$37:38:41 & 10:47:57.02 & $-$37:37:35.76  & 3621 & 0.022 & 2.49$\pm$0.20 & 0.04 & 1.00 & 49.22/67$^{*}$\\
 	J1117.2$-$5338 & 11:17:15 & $-$53:38:03 & 11:17:14.74 & $-$53:38:15.04  & 3361 & 0.012 & 1.97$\pm$0.33 & 0.15 & 0.85 & 40.52/37$^{*}$\\
 	J1125.0$-$5806 & 11:25:05 & $-$58:06:42 & 11:25:04.49 & $-$58:05:40.78  & 3094 & 0.019 & 2.49$\pm$0.24 & 0.37 & 2.15 & 58.74/55$^{*}$\\
 	J1145.9$-$0637 & 11:45:54 & $-$06:37:18 & 11:46:00.65 & $-$06:38:51.36  & 2343 & 0.024 & 1.98$\pm$0.24 & 0.03 & 3.01 & 54.65/48$^{*}$\\
 	J1220.1$-$2459 & 12:20:11 & $-$24:59:07 & 12:20:14.54 & $-$24:59:52.08  & 4233 & 0.061 & 2.00$\pm$0.12 & 0.08 & 3.07 & 9.56/10$^{*}$\\
 	J1220.4$-$3714 & 12:20:26 & $-$37:14:31 & 12:20:19.78 & $-$37:14:10.68  & 1354 & 0.051 & 2.35$\pm$0.20 & 0.08 & 2.43 & 39.28/54$^{*}$\\
 	J1234.8$-$0435 & 12:34:50 & $-$04:35:11 & 12:34:47.96 & $-$04:32:46.26  & 3424 & 0.007 & 2.15$\pm$0.29 & 0.03 & 0.29 & 22.67/23$^{*}$\\
 	J1240.5$-$7148 & 12:40:36 & $-$71:48:40 & 12:40:21.19 & $-$71:48:57.24  & 4695 & 0.272 & 1.84$\pm$0.04 & 0.14 & 14.98 & 78.43/55$^{*}$\\
 	J1248.8+5128 & 12:48:54 & +51:28:08 & 12:48:34.30 & +51:28:06.96  & 6832 & 0.010 & 1.83$\pm$0.20 & 0.01 & 0.47 & 17.25/21$^{*}$\\
 	J1249.2$-$2809 & 12:49:13 & $-$28:09:37 & 12:49:19.42 & $-$28:08:38.04  & 5956 & 0.101 & 2.17$\pm$0.07 & 0.06 & 4.30 & 25.68/27\\
 J1447.0$-$2657 & 14:47:04 & $-$26:57:50 & 14:46:57.41 & $-$26:57:01.80  & 4106 & 0.239 & 1.89$\pm$0.05 & 0.09 & 13.8 & 21.88/42\\
 	J1517.0+2638 & 15:17:02 & +26:38:31 & 15:16:49.82 & +26:36:42.48  & 3514 & 0.005 & 2.13$\pm$0.46 & 0.04 & 0.23 & 18.49/17$^{*}$\\
 	J1541.7+1413 & 15:41:44 & +14:13:48 & 15:41:49.85 & +14:14:44.52  & 2878 & 0.005 & 2.61$\pm$0.62 & 0.03 & 0.21 & 16.01/14$^{*}$  \\
 	J1553.8$-$2425 & 15:53:51 & $-$24:25:14 & 15:53:31.25 & $-$24:22:05.16  & 5384 & 0.013& 1.39$\pm$0.27 & 0.11 & 0.92 & 35.75/60\\
 	J1650.9+0430 & 16:51:00 & +04:30:05 & 16:50:58.87 & +04:27:34.92  & 14020 &0.001 & \nodata  & 0.06 & \nodata & \nodata $\dagger$\\
 	J1704.5$-$0527 & 17:04:34 & $-$05:27:25 & 17:04:34.10 & $-$05:28:37.92  & 6309 & 0.063& 1.88$\pm$0.09 & 0.13 & 3.56 & 24.32/18\\
 	J1856.1$-$1221 & 18:56:07 & $-$12:21:16 & 18:56:06.67 & $-$12:21:50.40  & 3286 & 0.023& 1.93$\pm$0.28 & 0.16 & 1.63 & 21.00/21$^{*}$\\
 	J1958.1+2437 & 19:58:07 & +24:37:51 & 19:58:00.38 & +24:38:03.84  & 3529 & 0.019& 2.39$\pm$0.30 & 0.69 & 3.13 & 59.42/55$^{*}$\\
 	J2036.0+4901 & 20:36:03 & +49:01:11 & 20:35:51.24 & +49:01:39.72  & 3521 & 0.004& 2.01$\pm$1.17 &0.66 & 0.13 &  16.01/11$^{*}$ \\
 	J2042.7+1520 & 20:42:45 & +15:20:03 & 20:42:59.71 & +15:21:06.48  & 17600 & 0.035& 2.23$\pm$0.08 & 0.07 & 1.89 & 29.92/26\\
 	J2109.7+0440 & 21:09:42 & +04:40:11 & 21:09:40.08 & +04:39:59.40  & 354 & 0.025 & 1.58$\pm$0.73 & 0.07 & 1.69 & 6.19/6$^{*}$\\
 	J2115.2+1218 & 21:15:15 & +12:18:29 & 21:15:22.10 & +12:18:02.88  & 3756 & 0.007& 2.95$\pm$0.36 & 0.05 & 0.37 & 21.28/25$^{*}$\\
 	J2142.3+3659 & 21:42:22 & +36:59:40 & 21:42:26.50 & +36:59:48.12  & 4365 & 0.022& 2.16$\pm$0.14 & 0.17 & 6.73 & 6.69/7\\
 	J2142.7+1959 & 21:42:44 & +19:59:22 & 21:42:47.45 & +19:58:09.84  & 1628 & 0.123& 2.99$\pm$0.22 & 0.07 & 1.88 & 64.00/69$^{*}$\\
 	J2151.5+4155 & 21:51:32 & +41:55:47 & 21:51:22.90 & +41:56:32.28  & 3284 & 0.022& 2.81$\pm$0.32 & 0.23 & 1.53 & 26.37/62\\
 	J2308.8+5424 & 23:08:54 & +54:24:22 & 23:08:48.62 & +54:26:08.88  & 1051 & 0.022& 2.24$\pm$0.53 & 0.28 & 2.11 & 18.41/20$^{*}$\\ 

 	&  &  &  &  &  &  &  & \\
 	\enddata
    \tablecomments{Sources with $\dagger$ sign did not yield a reliable fit with a simple powerlaw model.}
    \tablenotetext{*}{Cash statistics was employed for model fitting}
 	\tablenotetext{a,b}{3FHL Right Ascension, 3FHL Declination} 
    \tablenotetext{c,d}{$Swift$-XRT Right Ascension, $Swift$-XRT Declination}
\tablenotetext{e}{X-Ray Photon Index}
\tablenotetext{f}{Hydrogen Column Density ($\times$ 10$^{22}$)}
\tablenotetext{g}{Unabsorbed Flux in the 0.3-10 keV band ($\times$ 10$^{-12}$)}
 
\end{deluxetable}

 \begin{deluxetable}{lccccccccr}
 	\tabletypesize{\scriptsize}
 	\rotate
 	\tablecolumns{9}
 	\tablewidth{0pt}

 	\tablecaption{\small Cross-Correlation \label{tab:corr}}
 	\tablehead{
 		\colhead{3FHL\tablenotemark{1}} &
 		\colhead{1RXS\tablenotemark{2}} &
 		\colhead{Radio\tablenotemark{3}} &
 		\colhead{WISE\tablenotemark{4}} &
		 		\colhead{BZ Cat\tablenotemark{5}} &
	\colhead{MILLIQUAS\tablenotemark{6}} &
    	\colhead{2WHSP\tablenotemark{7}} &
 		\colhead{Class} &
 		\colhead{Redshift} &
 		 }
 	\startdata
 J0049.0+4224 & \nodata & N J004859+422350 & J004859.15+422351.1 &  \nodata &  \nodata   & 004859.0+422351  & \nodata&  \nodata  \\
 J0121.9$-$3917 & \nodata & S J012152$-$391542 & J012152.69$-$391544.2 &    \nodata   &  \nodata & \nodata & \nodata &\nodata  \\
 J0156.2$-$2419 & \nodata &  \nodata & J015624.54$-$242003.7   &  & J015624.6$-$242004  & \nodata & QSO 93\% \tablenotemark{c}  & \nodata\\
 J0213.9$-$6950 & \nodata &  \nodata & J021358.69$-$695137.0   &  \nodata &  \nodata  & \nodata & \nodata &  \nodata   \\
 J0251.2$-$1830 & \nodata & N J025111$-$183112 & J025111.52$-$183112.7   & \nodata  &  \nodata & \nodata &   \nodata &  \nodata    \\
 J0316.5$-$2610 & \nodata & N J031615$-$260755 & J031614.94$-$260757.2  & J0316$-$2607 & RXS J03162$-$2607 & 031614.8-260757	&      BL LAC \tablenotemark{a,b,c,d} & 0.443 \tablenotemark{a,c,d} \\
 J0350.4$-$5143 & \nodata &  \nodata & J035028.30$-$514454.3&  \nodata &  \nodata & \nodata & \nodata &  \nodata   \\
 J0350.8$-$2814 & \nodata &  \nodata & J035051.32$-$281632.8   &  \nodata &  \nodata & 035051.2-281632 & \nodata &   \nodata  \\
 J0401.0$-$5355 & \nodata &  \nodata & J040111.20$-$535458.5   &  \nodata & J040111.2$-$535458 & \nodata & QSO 86\% \tablenotemark{c}   &\nodata    \\
 J0427.5$-$6705 & \nodata &  \nodata & J042749.72$-$670434.7   &  \nodata & \nodata & \nodata &   \nodata  \\
 J0438.0$-$7328 & \nodata & S J043836$-$732921 & J043837.07$-$732921.6 & \nodata &   \nodata &\nodata & LEDA 255538   \tablenotemark{e} & \nodata \\
 J0506.9+0323 & \nodata & N J050650+032401 & J050650.14+032358.7   &  \nodata & \nodata & \nodata &    \nodata   \\
 J0541.1$-$4855 & \nodata & \nodata & J054106.92$-$485410.3   &  \nodata & J054106.9$-$485412 & \nodata& QSO 91\% & \nodata \tablenotemark{c}\\

 J0559.6+3045 & \nodata &  N J055941+304227  &  
 &  \nodata &  \nodata & \nodata & \nodata &   \nodata  \\

J0706.1+0247 & \nodata & N J070610+024449   &  \nodata &  \nodata &  \nodata & \nodata & \nodata &   \nodata  \\

 J0725.7$-$0548 & \nodata   &  \nodata &  \nodata &  \nodata &  \nodata & \nodata & \nodata &  \nodata   \\
 J0739.7$-$6720 &  J073928.1$-$672147(15) &  \nodata& 
 &   \nodata &  \nodata & \nodata & \nodata &   \nodata  \\
 J0747.7$-$4927 &  J074725.1$-$492626(12) &  \nodata &  J074724.74-492633.1 &  \nodata &   \nodata & 074724.7-492633 & \nodata &  \nodata   \\
 J0813.7$-$0353 & \nodata & N J081338$-$035716 & J081338.07$-$035716.7 &  \nodata &  \nodata  & \nodata & \nodata &  \nodata \\

 J0820.2$-$2803 & \nodata &  \nodata   &  \nodata &  \nodata &  \nodata & \nodata & \nodata &  \nodata   \\

J0928.5$-$5256 & \nodata &  \nodata & J092818.63$-$525703.1   &  \nodata &  \nodata & \nodata & \nodata &  \nodata  \\
 J0937.8$-$1434 & \nodata   &  \nodata & \nodata  &  \nodata &  \nodata & \nodata  & \nodata  &   \nodata  \\
 J1016.2$-$4245 &  J101620.6$-$424733(14) & \nodata  & J101620.67$-$424722.6   &  \nodata  &  \nodata  & \nodata & \nodata   &  \nodata    \\

 J1024.5$-$4543 & \nodata  &   \nodata & J102432.37$-$454426.9   &  \nodata  &  \nodata  & \nodata & \nodata  &  \nodata    \\

J1033.4$-$5033 &  J103332.0$-$503539(14) &  & J103332.15$-$503528.8   &   \nodata &  \nodata  & \nodata & \nodata  &   \nodata   \\

 J1034.8$-$4645 & \nodata  & \nodata & J103438.49$-$464403.5   &  \nodata  &  \nodata  & \nodata &  \nodata  & \nodata     \\
 J1047.9$-$3738 & \nodata & \nodata & J104756.94$-$373730.8   & \nodata   &  \nodata  & 104756.8-373730 & \nodata  &  \nodata    \\

 J1117.2$-$5338 &  \nodata   &  \nodata  &  \nodata  &  \nodata  &  \nodata  & \nodata & \nodata  &    \nodata  \\

 J1125.0$-$5806 &  J112502.3$-$580547(16) &  \nodata  & J112503.99$-$580539.9   &   \nodata &  \nodata & \nodata &  \nodata  &  \nodata    \\
 J1145.9$-$0637 &\nodata   &  \nodata  & J114600.85$-$063854.9   &  \nodata  &  \nodata  & \nodata & \nodata  &   \nodata   \\
 J1220.1$-$2459 &\nodata   & N J122014$-$245949 & J122014.53$-$245948.6   &  \nodata  & J122014.5$-$245948 & \nodata & \nodata &   \nodata     \\
 J1220.4$-$3714 & \nodata  & N J122020$-$371411 & J122019.81$-$371414.2  &  \nodata  &  \nodata  & 122019.8-371413 & \nodata  &  \nodata    \\
 J1234.8$-$0435 & \nodata  &  \nodata  & J123448.05$-$043245.2  &  \nodata  &  \nodata  & \nodata  & Galaxy\tablenotemark{f} &0.277 \tablenotemark{f}   \\

 J1240.5$-$7148 &  J124015.2$-$714859(21) &  \nodata  & J124021.21$-$714857.7   & \nodata   & \nodata  & \nodata & \nodata   & \nodata   \\
 J1248.8+5128 &  J124832.1+512817(12) & N J124834+512807 & J124834.29+512807.8  &   J1248+5128 &  SDSS J124834.30+512807.8 & 	 \nodata &      BL LAC\tablenotemark{a,d} & 0.351\tablenotemark{a,b,c,d} \\
 J1249.2$-$2809 &  \nodata   & N J124919$-$280833 & J124919.31$-$280834.4   &  \nodata  &  \nodata  &  124919.3-280834  & LEDA 745327 \tablenotemark{e} & \nodata \\
 J1447.0$-$2657 &   \nodata    &  \nodata  &  \nodata  &  \nodata  &  \nodata  & \nodata & \nodata  &  \nodata    \\
 J1517.0+2638 &   \nodata    &   \nodata &  \nodata  & \nodata   &   \nodata & \nodata & \nodata  &   \nodata   \\
 J1541.7+1413 &   \nodata    &  \nodata  & \nodata   & \nodata   &  \nodata  & \nodata & \nodata  &  \nodata    \\
 J1553.8$-$2425 &  \nodata   & N J155331.6$-$242206   & \nodata   &  \nodata  & PKS 1550$-$242 & \nodata & BL Lac\tablenotemark{a,d} & 0.332 \tablenotemark{a,b,c,d}   \\
 J1650.9+0430 &  \nodata   &   \nodata & J165058.69+042734.9   &  \nodata  &  \nodata  &\nodata & \nodata  &   \nodata  \\
 J1704.5$-$0527 & \nodata    &  \nodata  & J170433.83$-$052840.7  &  \nodata  & \nodata   & 170433.7-052840&  \nodata  &  \nodata    \\

 J1856.1$-$1221 &  \nodata   & N J185606$-$122148   &  \nodata  &  \nodata  &  \nodata  & \nodata &\nodata  &   \nodata   \\

 J1958.1+2437 &  \nodata   & N J195800+243802 & J195800.50+243800.9   &  \nodata  & \nodata   &\nodata  & \nodata  &   \nodata   \\

 J2036.0+4901 & \nodata  & N J203551+490143   &  \nodata  & \nodata   &   \nodata &  & \nodata  &  \nodata    \\
 J2042.7+1520 &  \nodata & N J204259+152107 & J204259.72+152108.1   &  \nodata  & J204259.7+152108 &  &   QSO 81 \tablenotemark{c}&  \nodata \\
 J2109.7+0440 & \nodata  & N J210939+044000 & J210940.12+044000.6   & \nodata   & SDSS J210940.12+044000.3 &    &   QSO TYPE 1\tablenotemark{c} &1.4$^{\dagger}$ \tablenotemark{c} \\
 J2115.2+1218 & \nodata  & N J211522+121802 & J211522.00+121802.6   &  \nodata  & SDSS J211522.00+121802.8 &  \nodata &  QSO TYPE 1 \tablenotemark{c} &2.1$^{\dagger}$ \tablenotemark{c}\\
 J2142.3+3659 &  \nodata    & N J214226+365949 & J214226.49+365949.7   &  \nodata  & \nodata   &214226.4+365948&  \nodata  &  \nodata    \\
 J2142.7+1959 &  \nodata   & N J214247+195810 & J214247.62+195810.9   &  \nodata  & \nodata   & \nodata &  \nodata  &   \nodata   \\

 J2151.5+4155 & \nodata  & N J215122+415632 & J215123.22+415633.9   &  \nodata  &  \nodata & \nodata & \nodata   &   \nodata   \\

 J2308.8+5424 & \nodata  & N J230848+542612 & J230848.74+542611.2  & \nodata  & \nodata & \nodata & \nodata  &  \nodata  \\
    \enddata
{
\tablecomments{The redshifts marked with $\dagger$ were derived using a clustering method based on the photometric SDSS data. See \citet{Flesch2015} for details.}
 \tablenotetext{1}{3FHL name \citep{Ajello2017}} 
 \tablenotetext{2}{1RXS name with positional uncertainty in arcsec\citep{Voges1999,Voges2000}}
  \tablenotetext{3}{NVSS/SUMSS name; N for NVSS and S for SUMSS \citep{Condon1998,Mauch2003}}
   \tablenotetext{4}{WISE name;\citep{Cutri2013}}
    \tablenotetext{5}{BZ Catalog name, BZB \citep{Massaro2015a}}
     \tablenotetext{6}{The Million Quasar Catalog \citep{Flesch2017}}
          \tablenotetext{7}{Wise High Synchrotron Peaked blazars Catalog \citep{Chang2016}}

    \tablerefs{$^{a}$DAbrusco2014, $^{b}$Flesch2015, $^{c}$Flesch2017, $^{d}$Massaro2015a, $^{e}$Paturel2003, $^{f}$Colless2001}

}
\end{deluxetable}
\clearpage
\begin{longtable}{|l|l|r|}
\caption[]{Parameters for Blazars for classification
\label{tab:pars}}\\
\hline
\hline 
Parameter & Catalog	& References\\ 
\hline 
\hline 
    X-ray Spectral Index  & Table~\ref{tab:swift} for unknown sample & See Table \ref{tab:swift}\\ 
    &  1SXPS for training set & \citet{Evans2014} \\
\hline 
    Variability Bayes Blocks  &  3FHL & \citet{Ajello2017}  \\
\hline 
     w1-w2  & AllWISE & \citet{Cutri2013}\\
\hline 
    w2-w3   & AllWISE & \citet{Cutri2013} \\
\hline 
   Gamma-ray Spectral Index 1 & 3FGL & \citet{Acero2015}\\
\hline 
   Gamma-ray Spectral Index 2 & 3FHL & \citet{Ajello2017}\\
\hline 
\hline\end{longtable}
\clearpage
\begin{small}
	\begin{longtable}{lrrrrrr}
		\caption[]{Machine Learning Results}\\
		\label{tab:mlcomp}\\
	\hline
	\multicolumn{1}{c}{3FHL} &
	\multicolumn{1}{c}{DT Pred\tablenotemark{1}} &
	\multicolumn{1}{c}{SVM Pred\tablenotemark{2}} &
	\multicolumn{1}{c}{SVM Prob\tablenotemark{3}} &
	\multicolumn{1}{c}{RF Pred\tablenotemark{4}} &
	\multicolumn{1}{c}{RF Prob\tablenotemark{5}} & \\
	\hline
    
 J0049.0+4224 & fsrq & bll & 1.0 & bll & 0.84\\
  J0121.9-3917 & bll & bll & 1.0 & bll & 0.85\\
  J0156.2-2419 & bll & bll & 0.94 & bll & 0.96\\
  J0213.9-6950 & bll & bll & 0.98 & bll & 0.69\\
  J0251.2-1830 & bll & bll & 0.95 & bll & 0.99\\
  J0316.5-2610 & bll & bll & 0.98 & bll & 0.99\\
  J0350.4-5143 & bll & bll & 0.98 & bll & 0.92\\
  J0350.8-2814 & bll & bll & 1.0 & bll & 0.85\\
  J0401.0-5355 & fsrq & bll & 0.99 & bll & 0.8\\
  J0427.5-6705 & fsrq & bll & 1.0 & bll & 0.81\\
  J0438.0-7328 & bll & bll & 1.0 & bll & 0.79\\
  J0506.9+0323 & bll & bll & 0.98 & bll & 0.95\\
  J0541.1-4855 & fsrq & bll & 1.0 & bll & 0.77\\
  J0739.7-6720 & bll & bll & 1.0 & bll & 0.74\\
  J0747.7-4927 & bll & bll & 1.0 & bll & 0.88\\
  J0813.7-0353 & bll & bll & 1.0 & bll & 0.85\\
  J0937.8-1434 & bll & bll & 1.0 & bll & 0.9\\
  J1016.2-4245 & bll & bll & 0.98 & bll & 1.0\\
  J1047.9-3738 & bll & bll & 0.98 & bll & 0.91\\
  J1145.9-0637 & bll & bll & 1.0 & bll & 0.84\\
  J1220.1-2459 & bll & bll & 0.99 & bll & 0.87\\
  J1220.4-3714 & bll & bll & 0.98 & bll & 1.0\\
  J1234.8-0435 & bll & bll & 0.57 & bll & 0.93\\
  J1248.8+5128 & bll & bll & 0.87 & bll & 0.95\\
  J1249.2-2809 & bll & bll & 1.0 & bll & 0.86\\
  J1447.0-2657 & bll & bll & 0.99 & bll & 0.73\\
  J1517.0+2638 & bll & bll & 1.0 & bll & 0.84\\
  J1541.7+1413 & bll & bll & 1.0 & bll & 0.86\\
  J1553.8-2425 & bll & bll & 0.98 & bll & 0.84\\
  J1650.9+0430 & fsrq & bll & 1.0 & bll & 0.77\\
  J1704.5-0527 & bll & bll & 0.95 & bll & 0.97\\
  J2042.7+1520 & bll & bll & 0.99 & bll & 0.84\\
  J2109.7+0440 & bll & bll & 0.99 & bll & 0.88\\
  J2115.2+1218 & bll & bll & 0.89 & bll & 0.97\\
  J2142.3+3659 & bll & bll & 1.0 & bll & 0.84\\
  J2142.7+1959 & bll & bll & 0.99 & bll & 0.87\\
\hline
\end{longtable}
\begin{samepage}
\begin{itemize}
  \item[1]{Predicted Class by the Decision Tree Algorithm}
    \item[2]{Predicted Class by the SVM Algorithm}
  \item[3]{Probability to be a BLL associated by the SVM Method}
  \item[4]{Predicted Class by the Random Forest (RF) Algorithm}
  \item[5]{Probability to be a BLL associated by the RF Method}
  \end{itemize}
\end{samepage}

\end{small}


\acknowledgments
This publication made use of data products from the Wide-field Infrared Survey Explorer, which is a joint project of the University of California, Los Angeles, and the Jet Propulsion Laboratory/California Institute of Technology, funded by the National Aeronautics and Space Administration. This work also utilized the data supplied by UK $Swift$ Data Centre at the University of Leicester as well as of TOPCAT software\citep{Taylor2005}. This research has also made use of the SIMBAD database,
operated at CDS, Strasbourg, France; and the NASA/IPAC Extragalactic Database (NED) which is operated by the Jet Propulsion Laboratory, California Institute of Technology, under contract with the National Aeronautics and Space Administration. This research has made use of data and/or software provided by the High Energy Astrophysics Science Archive Research Center (HEASARC), which is a service of the Astrophysics Science Division at NASA/GSFC and the High Energy Astrophysics Division of the Smithsonian Astrophysical Observatory. \\
 We are grateful to Eric D. Feigelson for discussions and suggestions regarding the correction for the imbalanced classes part for the machine learning methods.

\clearpage

\bibliographystyle{apj}
\bibliography{bibliography}

\begin{thebibliography}{}
\expandafter\ifx\csname natexlab\endcsname\relax\def\natexlab#1{#1}\fi

\bibitem[{Abdo {et~al.}(2010)Abdo, Ackermann, Agudo, Ajello, Aller, Aller,
  Angelakis, Arkharov, Axelsson, Bach, Baldini, Ballet, Barbiellini, Bastieri,
  Baughman, Bechtol, Bellazzini, Benitez, Berdyugin, Berenji, Blandford, Bloom,
  Boettcher, Bonamente, Borgland, Bregeon, Brez, Brigida, Bruel, Burnett,
  Burrows, Buson, Caliandro, Calzoletti, Cameron, Capalbi, Caraveo, Carosati,
  Casandjian, Cavazzuti, Cecchi, {\c{C}}elik, Charles, Chaty, Chekhtman, Chen,
  Chiang, Chincarini, Ciprini, Claus, Cohen-Tanugi, Colafrancesco, Cominsky,
  Conrad, Costamante, Cutini, D'ammando, Deitrick, D'Elia, Dermer, de~Angelis,
  de~Palma, Digel, Donnarumma, Silva, Drell, Dubois, Dultzin, Dumora, Falcone,
  Farnier, Favuzzi, Fegan, Focke, Forn{\'{e}}, Fortin, Frailis, Fuhrmann,
  Fukazawa, Funk, Fusco, G{\'{o}}mez, Gargano, Gasparrini, Gehrels, Germani,
  Giebels, Giglietto, Giommi, Giordano, Giuliani, Glanzman, Godfrey, Grenier,
  Gronwall, Grove, Guillemot, Guiriec, Gurwell, Hadasch, Hanabata, Harding,
  Hayashida, Hays, Healey, Heidt, Hiriart, Horan, Hoversten, Hughes, Itoh,
  Jackson, J{\'{o}}hannesson, Johnson, Johnson, Jorstad, Kadler, Kamae,
  Katagiri, Kataoka, Kawai, Kennea, Kerr, Kimeridze, Kn{\"{o}}dlseder, Kocian,
  Kopatskaya, Koptelova, Konstantinova, Kovalev, Kovalev, Kurtanidze, Kuss,
  Lande, Larionov, Latronico, Leto, Lindfors, Longo, Loparco, Lott, Lovellette,
  Lubrano, Madejski, Makeev, Marchegiani, Marscher, Marshall, Max-Moerbeck,
  Mazziotta, McConville, McEnery, Meurer, Michelson, Mitthumsiri, Mizuno,
  Moiseev, Monte, Monzani, Morselli, Moskalenko, Murgia, Nestoras, Nilsson,
  Nizhelsky, Nolan, Norris, Nuss, Ohsugi, Ojha, Omodei, Orlando, Ormes,
  Osborne, Ozaki, Pacciani, Padovani, Pagani, Page, Paneque, Panetta, Parent,
  Pasanen, Pavlidou, Pelassa, Pepe, Perri, Pesce-Rollins, Piranomonte, Piron,
  Pittori, Porter, Puccetti, Rahoui, Rain{\`{o}}, Raiteri, Rando, Razzano,
  Reimer, Reimer, Reposeur, Richards, Ritz, Rochester, Rodriguez, Romani, Ros,
  Roth, Roustazadeh, Ryde, Sadrozinski, Sadun, Sanchez, Sander, Parkinson,
  Scargle, Sellerholm, Sgr{\`{o}}, Shaw, Sigua, Siskind, Smith, Smith, Spandre,
  Spinelli, Starck, Stevenson, Stratta, Strickman, Suson, Tajima, Takahashi,
  Takahashi, Takalo, Tanaka, Thayer, Thayer, Thompson, Tibaldo, Torres, Tosti,
  Tramacere, Uchiyama, Usher, Vasileiou, Verrecchia, Vilchez, Villata, Vitale,
  Waite, Wang, Winer, Wood, Ylinen, Zensus, Zhekanis, \& Ziegler}]{Abdo2010b}
Abdo, A.~A., Ackermann, M., Agudo, I., {et~al.} 2010, The Astrophysical
  Journal, 716, 30

\bibitem[{Acero {et~al.}(2015)Acero, Ackermann, Ajello, Albert, Atwood,
  Axelsson, Baldini, Ballet, Barbiellini, Bastieri, Belfiore, Bellazzini,
  Bissaldi, Blandford, Bloom, Bogart, Bonino, Bottacini, Bregeon, Britto,
  Bruel, Buehler, Burnett, Buson, Caliandro, Cameron, Caputo, Caragiulo,
  Caraveo, Casandjian, Cavazzuti, Charles, Chaves, Chekhtman, Cheung, Chiang,
  Chiaro, Ciprini, Claus, Tanugi, Cominsky, Conrad, Cutini, D'Ammando,
  de~Angelis, DeKlotz, de~Palma, Desiante, Digel, Venere, Drell, Dubois,
  Dumora, Favuzzi, Fegan, Ferrara, Finke, Franckowiak, Fukazawa, Funk, Fusco,
  Gargano, Gasparrini, Giebels, Giglietto, Giommi, Giordano, Giroletti,
  Glanzman, Godfrey, Grenier, Grondin, Grove, Guillemot, Guiriec, Hadasch,
  Harding, Hays, Hewitt, Hill, Horan, Iafrate, Jogler, J{\'{o}}hannesson,
  Johnson, Johnson, Johnson, Johnson, Kamae, Kataoka, Katsuta, Kuss, Mura,
  Landriu, Larsson, Latronico, Goumard, Li, Li, Longo, Loparco, Lott,
  Lovellette, Lubrano, Madejski, Massaro, Mayer, Mazziotta, McEnery, Michelson,
  Mirabal, Mizuno, Moiseev, Mongelli, Monzani, Morselli, Moskalenko, Murgia,
  Nuss, Ohno, Ohsugi, Omodei, Orienti, Orlando, Ormes, Paneque, Panetta,
  Perkins, Rollins, Piron, Pivato, Porter, Racusin, Rando, Razzano, Razzaque,
  Reimer, Reimer, Reposeur, Rochester, Romani, Salvetti, Conde, Parkinson,
  Schulz, Siskind, Smith, Spada, Spandre, Spinelli, Stephens, Strong, Suson,
  Takahashi, Takahashi, Tanaka, Thayer, Thayer, Thompson, Tibaldo, Tibolla,
  Torres, Torresi, Tosti, Troja, Klaveren, Vianello, Winer, Wood, Wood, \&
  Zimmer}]{Acero2015}
Acero, F., Ackermann, M., Ajello, M., {et~al.} 2015, The Astrophysical Journal
  Supplement Series, 218, 23

\bibitem[{Ackermann {et~al.}(2012)Ackermann, Ajello, Allafort, Antolini,
  Baldini, Ballet, Barbiellini, Bastieri, Bellazzini, Berenji, Blandford,
  Bloom, Bonamente, Borgland, Bouvier, Brandt, Bregeon, Brigida, Bruel,
  Buehler, Burnett, Buson, Caliandro, Cameron, Caraveo, Casandjian, Cavazzuti,
  Cecchi, {\c{C}}elik, Charles, Chekhtman, Chen, Cheung, Chiang, Ciprini,
  Claus, Cohen-Tanugi, Conrad, Cutini, {De Angelis}, Decesar, {De Luca}, {De
  Palma}, Dermer, {Do Couto E Silva}, Drell, Drlica-Wagner, Dubois, Enoto,
  Favuzzi, Fegan, Ferrara, Focke, Fortin, Fukazawa, Funk, Fusco, Gargano,
  Gasparrini, Gehrels, Germani, Giglietto, Giordano, Giroletti, Glanzman,
  Godfrey, Grenier, Grondin, Grove, Guillemot, Guiriec, Gustafsson, Hadasch,
  Hanabata, Harding, Hayashida, Hays, Healey, Hill, Horan, Hou,
  J{\'{o}}hannesson, Johnson, Johnson, Kamae, Katagiri, Kataoka, Kerr,
  Kn{\"{o}}dlseder, Kuss, Lande, Latronico, Lee, Lemoine-Goumard, Longo,
  Loparco, Lott, Lovellette, Lubrano, Madejski, Mazziotta, McEnery, Mehault,
  Michelson, Mignani, Mitthumsiri, Mizuno, Monte, Monzani, Morselli,
  Moskalenko, Murgia, Nakamori, Naumann-Godo, Nolan, Norris, Nuss, Ohsugi,
  Okumura, Omodei, Orlando, Ormes, Ozaki, Paneque, Panetta, Parent, Pelassa,
  Pesce-Rollins, Pierbattista, Piron, Pivato, Porter, Rain{\`{o}}, Rando, Ray,
  Razzano, Reimer, Reimer, Reposeur, Romani, Sadrozinski, Salvetti, Parkinson,
  Schalk, Sgr{\`{o}}, Shaw, Siskind, Smith, Spandre, Spinelli, Suson,
  Takahashi, Tanaka, Thayer, Thayer, Thompson, Tibaldo, Tibolla, Torres, Tosti,
  Tramacere, Troja, Usher, Vandenbroucke, Vasileiou, Vianello, Vilchez, Vitale,
  Waite, Wallace, Wang, Winer, Wolff, Wood, Wood, Yang, \&
  Zimmer}]{Ackermann2012}
Ackermann, M., Ajello, M., Allafort, A., {et~al.} 2012, Astrophysical Journal,
  753, arXiv:1108.1202

\bibitem[{Ackermann {et~al.}(2013)Ackermann, Ajello, Allafort, Atwood, Baldini,
  Ballet, Barbiellini, Bastieri, Bechtol, Belfiore, Bellazzini, Bernieri,
  Bissaldi, Bloom, Bonamente, Brandt, Bregeon, Brigida, Bruel, Buehler,
  Burnett, Buson, Caliandro, Cameron, Campana, Caraveo, Casandjian, Cavazzuti,
  Cecchi, Charles, Chaves, Chekhtman, Cheung, Chiang, Chiaro, Ciprini, Claus,
  Cohen-Tanugi, Cominsky, Conrad, Cutini, D'Ammando, de~Angelis, de~Palma,
  Dermer, Desiante, Digel, {Di Venere}, Drell, Drlica-Wagner, Favuzzi, Fegan,
  Ferrara, Focke, Fortin, Franckowiak, Funk, Fusco, Gargano, Gasparrini,
  Gehrels, Germani, Giglietto, Giommi, Giordano, Giroletti, Godfrey,
  Gomez-Vargas, Grenier, Guiriec, Hadasch, Hanabata, Harding, Hayashida, Hays,
  Hewitt, Hill, Horan, Hughes, Jogler, J{\'{o}}hannesson, Johnson, Johnson,
  Johnson, Kamae, Kataoka, Kawano, Kn{\"{o}}dlseder, Kuss, Lande, Larsson,
  Latronico, Lemoine-Goumard, Longo, Loparco, Lott, Lovellette, Lubrano,
  Massaro, Mayer, Mazziotta, McEnery, Mehault, Michelson, Mizuno, Moiseev,
  Monzani, Morselli, Moskalenko, Murgia, Nemmen, Nuss, Ohsugi, Okumura,
  Orienti, Ormes, Paneque, Perkins, Pesce-Rollins, Piron, Pivato, Porter,
  Rain{\`{o}}, Razzano, Reimer, Reimer, Reposeur, Ritz, Romani, Roth, {Saz
  Parkinson}, Schulz, Sgr{\`{o}}, Siskind, Smith, Spandre, Spinelli, Stawarz,
  Strong, Suson, Takahashi, Thayer, Thayer, Thompson, Tibaldo, Tinivella,
  Torres, Tosti, Troja, Uchiyama, Usher, Vandenbroucke, Vasileiou, Vianello,
  Vitale, Werner, Winer, Wood, \& Wood}]{Ackermann2013a}
---. 2013, The Astrophysical Journal Supplement Series, 209, 34

\bibitem[{Ackermann {et~al.}(2015)Ackermann, Ajello, Atwood, Baldini, Ballet,
  Barbiellini, Bastieri, Gonzalez, Bellazzini, Bissaldi, Blandford, Bloom,
  Bonino, Bottacini, Brandt, Bregeon, Britto, Bruel, Buehler, Buson, Caliandro,
  Cameron, Caragiulo, Caraveo, Carpenter, Casandjian, Cavazzuti, Cecchi,
  Charles, Chekhtman, Cheung, Chiang, Chiaro, Ciprini, Claus, Cohen-Tanugi,
  Cominsky, Conrad, Cutini, D'Abrusco, D'Ammando, de~Angelis, Desiante, Digel,
  Venere, Drell, Favuzzi, Fegan, Ferrara, Finke, Focke, Franckowiak, Fuhrmann,
  Fukazawa, Furniss, Fusco, Gargano, Gasparrini, Giglietto, Giommi, Giordano,
  Giroletti, Glanzman, Godfrey, Grenier, Grove, Guiriec, Hewitt, Hill, Horan,
  Itoh, J{\'{o}}hannesson, Johnson, Johnson, Kataoka, Kawano, Krauss, Kuss,
  Mura, Larsson, Latronico, Leto, Li, Li, Longo, Loparco, Lott, Lovellette,
  Lubrano, Madejski, Mayer, Mazziotta, McEnery, Michelson, Mizuno, Moiseev,
  Monzani, Morselli, Moskalenko, Murgia, Nuss, Ohno, Ohsugi, Ojha, Omodei,
  Orienti, Orlando, Paggi, Paneque, Perkins, Pesce-Rollins, Piron, Pivato,
  Porter, Rain{\`{o}}, Rando, Razzano, Razzaque, Reimer, Reimer, Romani,
  Salvetti, Schaal, Schinzel, Schulz, Sgr{\`{o}}, Siskind, Sokolovsky, Spada,
  Spandre, Spinelli, Stawarz, Suson, Takahashi, Takahashi, Tanaka, Thayer,
  Thayer, Tibaldo, Torres, Torresi, Tosti, Troja, Uchiyama, Vianello, Winer,
  Wood, Zimmer, D'Abrusco, D'Ammando, de~Angelis, Desiante, Digel, Venere,
  Drell, Favuzzi, Fegan, Ferrara, Finke, Focke, Franckowiak, Fuhrmann, Furniss,
  Fusco, Gargano, Gasparrini, Giglietto, Giommi, Giordano, Giroletti, Glanzman,
  Godfrey, Grenier, Grove, Guiriec, Hewitt, Hill, Horan, J'ohannesson, Johnson,
  Johnson, Kataoka, Kuss, Mura, Larsson, Latronico, Leto, Li, Li, Longo,
  Loparco, Lott, Lovellette, Lubrano, Madejski, Mayer, Mazziotta, McEnery,
  Michelson, Mizuno, Moiseev, Monzani, Morselli, Moskalenko, Murgia, Nuss,
  Ohno, Ohsugi, Ojha, Omodei, Orienti, Orlando, Paggi, Paneque, Perkins,
  Pesce-Rollins, Piron, Pivato, Porter, Rain`o, Rando, Razzano, Razzaque,
  Reimer, Reimer, Romani, Salvetti, Schaal, Schinzel, Schulz, Sgr`o, Siskind,
  Sokolovsky, Spada, Spandre, Spinelli, Stawarz, Suson, Takahashi, Takahashi,
  Tanaka, Thayer, Tibaldo, Torres, Torresi, Tosti, Troja, Uchiyama, Vianello,
  Winer, Wood, \& Zimmer}]{Ackermann2015b}
Ackermann, M., Ajello, M., Atwood, W.~B., {et~al.} 2015, The Astrophysical
  Journal, 810, 14

\bibitem[{Ackermann {et~al.}(2016{\natexlab{a}})Ackermann, Ajello, Atwood,
  Baldini, Ballet, Barbiellini, Bastieri, Gonzalez, Bellazzini, Bissaldi,
  Blandford, Bloom, Bonino, Bottacini, Brandt, Bregeon, Bruel, Buehler, Buson,
  Caliandro, Cameron, Caputo, Caragiulo, Caraveo, Cavazzuti, Cecchi, Charles,
  Chekhtman, Cheung, Chiang, Chiaro, Ciprini, Cohen, Cohen-Tanugi, Cominsky,
  Conrad, Cuoco, Cutini, D'Ammando, de~Angelis, de~Palma, Desiante, Mauro,
  Venere, Domínguez, Drell, Favuzzi, Fegan, Ferrara, Focke, Fortin,
  Franckowiak, Fukazawa, Funk, Furniss, Fusco, Gargano, Gasparrini, Giglietto,
  Giommi, Giordano, Giroletti, Glanzman, Godfrey, Grenier, Grondin, Guillemot,
  Guiriec, Harding, Hays, Hewitt, Hill, Horan, Iafrate, Hartmann, Jogler,
  J{\'{o}}hannesson, Johnson, Kamae, Kataoka, Kn{\"{o}}dlseder, Kuss, Mura,
  Larsson, Latronico, Lemoine-Goumard, Li, Li, Longo, Loparco, Lott,
  Lovellette, Lubrano, Madejski, Maldera, Manfreda, Mayer, Mazziotta,
  Michelson, Mirabal, Mitthumsiri, Mizuno, Moiseev, Monzani, Morselli,
  Moskalenko, Murgia, Nuss, Ohsugi, Omodei, Orienti, Orlando, Ormes, Paneque,
  Perkins, Pesce-Rollins, Petrosian, Piron, Pivato, Porter, Rain{\`{o}}, Rando,
  Razzano, Razzaque, Reimer, Reimer, Reposeur, Romani, S{\'{a}}nchez-Conde,
  Parkinson, Schmid, Schulz, Sgr{\`{o}}, Siskind, Spada, Spandre, Spinelli,
  Suson, Tajima, Takahashi, Takahashi, Takahashi, Thayer, Thompson, Tibaldo,
  Torres, Tosti, Troja, Vianello, Wood, Wood, Yassine, Zaharijas, \&
  Zimmer}]{Ackermann2016b}
---. 2016{\natexlab{a}}, The Astrophysical Journal Supplement Series, 222, 5

\bibitem[{Ackermann {et~al.}(2016{\natexlab{b}})Ackermann, Ajello, Atwood,
  Baldini, Ballet, Barbiellini, Bastieri, Gonzalez, Bellazzini, Bissaldi,
  Blandford, Bloom, Bonino, Bottacini, Brandt, Bregeon, Bruel, Buehler, Buson,
  Caliandro, Cameron, Caputo, Caragiulo, Caraveo, Cavazzuti, Cecchi, Charles,
  Chekhtman, Cheung, Chiang, Chiaro, Ciprini, Cohen, Cohen-Tanugi, Cominsky,
  Conrad, Cuoco, Cutini, D'Ammando, de~Angelis, de~Palma, Desiante, Mauro,
  Venere, Domínguez, Drell, Favuzzi, Fegan, Ferrara, Focke, Fortin,
  Franckowiak, Fukazawa, Funk, Furniss, Fusco, Gargano, Gasparrini, Giglietto,
  Giommi, Giordano, Giroletti, Glanzman, Godfrey, Grenier, Grondin, Guillemot,
  Guiriec, Harding, Hays, Hewitt, Hill, Horan, Iafrate, Hartmann, Jogler,
  J{\'{o}}hannesson, Johnson, Kamae, Kataoka, Kn{\"{o}}dlseder, Kuss, Mura,
  Larsson, Latronico, Lemoine-Goumard, Li, Li, Longo, Loparco, Lott,
  Lovellette, Lubrano, Madejski, Maldera, Manfreda, Mayer, Mazziotta,
  Michelson, Mirabal, Mitthumsiri, Mizuno, Moiseev, Monzani, Morselli,
  Moskalenko, Murgia, Nuss, Ohsugi, Omodei, Orienti, Orlando, Ormes, Paneque,
  Perkins, Pesce-Rollins, Petrosian, Piron, Pivato, Porter, Rain{\`{o}}, Rando,
  Razzano, Razzaque, Reimer, Reimer, Reposeur, Romani, S{\'{a}}nchez-Conde,
  Parkinson, Schmid, Schulz, Sgr{\`{o}}, Siskind, Spada, Spandre, Spinelli,
  Suson, Tajima, Takahashi, Takahashi, Takahashi, Thayer, Thompson, Tibaldo,
  Torres, Tosti, Troja, Vianello, Wood, Wood, Yassine, Zaharijas, \&
  Zimmer}]{Ackermann2016a}
---. 2016{\natexlab{b}}, The Astrophysical Journal Supplement Series, 222, 5

\bibitem[{Ajello {et~al.}(2015)Ajello, Gasparrini, Sanchez-Conde, Zaharijas,
  Gustafsson, Cohen-Tanugi, Dermer, Inoue, Hartmann, Ackermann, Bechtol,
  Franckowiak, Reimer, Romani, \& Strong}]{Ajello2015}
Ajello, M., Gasparrini, D., Sanchez-Conde, M., {et~al.} 2015, The Astrophysical
  Journal Letters, Volume 800, Issue 2, article id. L27, 7 pp. (2015)., 800,
  arXiv:1501.05301

\bibitem[{Ajello {et~al.}(2017)Ajello, Atwood, Baldini, Ballet, Barbiellini,
  Bastieri, Bellazzini, Bissaldi, Blandford, Bloom, Bonino, Bregeon, Britto,
  Bruel, Buehler, Buson, Cameron, Caputo, Caragiulo, Caraveo, Cavazzuti,
  Cecchi, Charles, Chekhtman, Cheung, Chiaro, Ciprini, Cohen, Costantin,
  Costanza, Cuoco, Cutini, D'Ammando, de~Palma, Desiante, Digel, {Di Lalla},
  {Di Mauro}, {Di Venere}, Dom{\'{i}}nguez, Drell, Dumora, Favuzzi, Fegan,
  Ferrara, Fortin, Franckowiak, Fukazawa, Funk, Fusco, Gargano, Gasparrini,
  Giglietto, Giommi, Giordano, Giroletti, Glanzman, Green, Grenier, Grondin,
  Grove, Guillemot, Guiriec, Harding, Hays, Hewitt, Horan, J{\'{o}}hannesson,
  Kensei, Kuss, {La Mura}, Larsson, Latronico, Lemoine-Goumard, Li, Longo,
  Loparco, Lott, Lubrano, Magill, Maldera, Manfreda, Mazziotta, McEnery, Meyer,
  Michelson, Mirabal, Mitthumsiri, Mizuno, Moiseev, Monzani, Morselli,
  Moskalenko, Negro, Nuss, Ohsugi, Omodei, Orienti, Orlando, Palatiello,
  Paliya, Paneque, Perkins, Persic, Pesce-Rollins, Piron, Porter, Principe,
  Rain{\`{o}}, Rando, Razzano, Razzaque, Reimer, Reimer, Reposeur, {Saz
  Parkinson}, Sgr{\`{o}}, Simone, Siskind, Spada, Spandre, Spinelli, Stawarz,
  Suson, Takahashi, Tak, Thayer, Thayer, Thompson, Torres, Torresi, Troja,
  Vianello, Wood, \& Wood}]{Ajello2017}
Ajello, M., Atwood, W.~B., Baldini, L., {et~al.} 2017, The Astrophysical
  Journal Supplement Series, 232, 18

\bibitem[{Alam {et~al.}(2015)Alam, Albareti, Prieto, Anders, Anderson, Andrews,
  Armengaud, Aubourg, Bailey, Bautista, Beaton, Beers, Bender, Berlind,
  Beutler, Bhardwaj, Bird, Bizyaev, Blake, Blanton, Blomqvist, Bochanski,
  Bolton, Bovy, Bradley, Brandt, Brauer, Brinkmann, Brown, Brownstein, Burden,
  Burtin, Busca, Cai, Capozzi, Rosell, Carrera, Chen, Chiappini, Chojnowski,
  Chuang, Clerc, Comparat, Covey, Croft, Cuesta, Cunha, da~Costa, {Da Rio},
  Davenport, Dawson, {De Lee}, Delubac, Deshpande, Dutra-Ferreira, Dwelly,
  Ealet, Ebelke, Edmondson, Eisenstein, Escoffier, Esposito, Fan,
  Fern{\'{a}}ndez-Alvar, Feuillet, Ak, Finley, Finoguenov, Flaherty, Fleming,
  Font-Ribera, Foster, Frinchaboy, Galbraith-Frew,
  Garc{\'{i}}a-Hern{\'{a}}ndez, P{\'{e}}rez, Gaulme, Ge, G{\'{e}}nova-Santos,
  Ghezzi, Gillespie, Girardi, Goddard, Gontcho, Hern{\'{a}}ndez, Grebel, Grieb,
  Grieves, Gunn, Guo, Harding, Hasselquist, Hawley, Hayden, Hearty, Ho, Hogg,
  Holley-Bockelmann, Holtzman, Honscheid, Huehnerhoff, Jiang, Johnson,
  Kinemuchi, Kirkby, Kitaura, Klaene, Kneib, Koenig, Lam, Lan, Lang, Laurent,
  Goff, Leauthaud, Lee, Lee, Licquia, Liu, Long, L{\'{o}}pez-Corredoira,
  Lorenzo-Oliveira, Lucatello, Lundgren, Lupton, Mack, Mahadevan, Maia,
  Majewski, Malanushenko, Malanushenko, Manchado, Manera, Mao, Maraston,
  Marchwinski, Margala, Martell, Martig, Masters, McBride, McGehee, McGreer,
  McMahon, M{\'{e}}nard, Menzel, Merloni, M{\'{e}}sz{\'{a}}ros, Miller,
  Miralda-Escud{\'{e}}, Miyatake, Montero-Dorta, More, Morice-Atkinson,
  Morrison, Muna, Myers, Newman, Neyrinck, Nguyen, Nichol, Nidever, Noterdaeme,
  Nuza, O'Connell, O'Connell, O'Connell, Ogando, Olmstead, Oravetz, Oravetz,
  Osumi, Owen, Padgett, Padmanabhan, Paegert, Palanque-Delabrouille, Pan,
  Parejko, Park, P{\^{a}}ris, Pattarakijwanich, Pellejero-Ibanez, Pepper,
  Percival, P{\'{e}}rez-Fournon, P{\'{e}}rez-R{\`{a}}fols, Petitjean, Pieri,
  Pinsonneault, de~Mello, Prada, Prakash, Price-Whelan, Raddick, Rahman, Reid,
  Rich, Rix, Robin, Rockosi, Rodrigues, Rodr{\'{i}}guez-Rottes, Roe, Ross,
  Ross, Rossi, Ruan, Rubi{\~{n}}o-Mart{\'{i}}n, Rykoff, Salazar-Albornoz,
  Salvato, Samushia, S{\'{a}}nchez, Santiago, Sayres, Schiavon, Schlegel,
  Schmidt, Schneider, Schultheis, Schwope, Sc{\'{o}}ccola, Sellgren, Seo,
  Shane, Shen, Shetrone, Shu, Sivarani, Skrutskie, Slosar, Smith, Sobreira,
  Stassun, Steinmetz, Strauss, Streblyanska, Swanson, Tan, Tayar, Terrien,
  Thakar, Thomas, Thompson, Tinker, Tojeiro, Troup, Vargas-Maga{\~{n}}a,
  Vazquez, Verde, Viel, Vogt, Wake, Wang, Weaver, Weinberg, Weiner, White,
  Wilson, Wisniewski, Wood-Vasey, Y{\`{e}}che, York, Zakamska, Zamora,
  Zasowski, Zehavi, Zhao, Zheng, Zhou, Zhou, Zhu, Zou, Thompson, Tinker,
  Tojeiro, Troup, Vargas-Maga{\~{n}}a, Vazquez, Verde, Viel, Vogt, Wake, Wang,
  Weaver, Weinberg, Weiner, White, Wilson, Wisniewski, Wood-Vasey, Ye`che,
  York, Zakamska, Zamora, Zasowski, Zehavi, Zhao, Zheng, Zhou, Zhou, Zou, \&
  Zhu}]{Alam2015}
Alam, S., Albareti, F.~D., Prieto, C.~A., {et~al.} 2015, The Astrophysical
  Journal Supplement Series, Volume 219, Issue 1, article id. 12, 27 pp.
  (2015)., 219, arXiv:1501.00963

\bibitem[{Arnaud(1996)}]{Arnaud1996}
Arnaud, K.~A. 1996, Astronomical Data Analysis Software and Systems V, 101, 17

\bibitem[{Atwood {et~al.}(2009)Atwood, Abdo, Ackermann, Althouse, Anderson,
  Axelsson, Baldini, Ballet, Band, Barbiellini, Bartelt, Bastieri, Baughman,
  Bechtol, B{\'{e}}d{\'{e}}r{\`{e}}de, Bellardi, Bellazzini, Berenji, Bignami,
  Bisello, Bissaldi, Blandford, Bloom, Bogart, Bonamente, Bonnell, Borgland,
  Bouvier, Bregeon, Brez, Brigida, Bruel, Burnett, Busetto, Caliandro, Cameron,
  Caraveo, Carius, Carlson, Casandjian, Cavazzuti, Ceccanti, Cecchi, Charles,
  Chekhtman, Cheung, Chiang, Chipaux, Cillis, Ciprini, Claus, Cohen-Tanugi,
  Condamoor, Conrad, Corbet, Corucci, Costamante, Cutini, Davis, Decotigny,
  Deklotz, Dermer, {De Angelis}, Digel, {Do Couto E Silva}, Drell, Dubois,
  Dumora, Edmonds, Fabiani, Farnier, Favuzzi, Flath, Fleury, Focke, Funk,
  Fusco, Gargano, Gasparrini, Gehrels, Gentit, Germani, Giebels, Giglietto,
  Giommi, Giordano, Glanzman, Godfrey, Grenier, Grondin, Grove, Guillemot,
  Guiriec, Haller, Harding, Hart, Hays, Healey, Hirayama, Hjalmarsdotter, Horn,
  Hughes, J{\'{o}}hannesson, Johansson, Johnson, Johnson, Johnson, Johnson,
  Kamae, Katagiri, Kataoka, Kavelaars, Kawai, Kelly, Kerr, Klamra,
  Kn{\"{o}}dlseder, Kocian, Komin, Kuehn, Kuss, Landriu, Latronico, Lee, Lee,
  Lemoine-Goumard, Lionetto, Longo, Loparco, Lott, Lovellette, Lubrano,
  Madejski, Makeev, Marangelli, Massai, Mazziotta, McEnery, Menon, Meurer,
  Michelson, Minuti, Mirizzi, Mitthumsiri, Mizuno, Moiseev, Monte, Monzani,
  Moretti, Morselli, Moskalenko, Murgia, Nakamori, Nishino, Nolan, Norris,
  Nuss, Ohno, Ohsugi, Omodei, Orlando, Ormes, Paccagnella, Paneque, Panetta,
  Parent, Pearce, Pepe, Perazzo, Pesce-Rollins, Picozza, Pieri, Pinchera,
  Piron, Porter, Poupard, Rain{\`{o}}, Rando, Rapposelli, Razzano, Reimer,
  Reimer, Reposeur, Reyes, Ritz, Rochester, Rodriguez, Romani, Roth, Russell,
  Ryde, Sabatini, Sadrozinski, Sanchez, Sander, Sapozhnikov, Parkinson,
  Scargle, Schalk, Scolieri, Sgr{\`{o}}, Share, Shaw, Shimokawabe, Shrader,
  Sierpowska-Bartosik, Siskind, Smith, Smith, Spandre, Spinelli, Starck,
  Stephens, Strickman, Strong, Suson, Tajima, Takahashi, Takahashi, Tanaka,
  Tenze, Tether, Thayer, Thayer, Thompson, Tibaldo, Tibolla, Torres, Tosti,
  Tramacere, Turri, Usher, Vilchez, Vitale, Wang, Watters, Winer, Wood, Ylinen,
  \& Ziegler}]{Atwood2009}
Atwood, W.~B., Abdo, A.~A., Ackermann, M., {et~al.} 2009, Astrophysical
  Journal, 697, 1071

\bibitem[{Breiman(2001)}]{Breiman2001}
Breiman, L. 2001, Machine Learning,
  arXiv:/dx.doi.org/10.1023{\%}2FA{\%}3A1010933404324

\bibitem[{Chang {et~al.}(2016)Chang, Arsioli, Giommi, \& Padovani}]{Chang2016}
Chang, Y.-L., Arsioli, B., Giommi, P., \& Padovani, P. 2016, Astronomy {\&}
  Astrophysics, 598, A17

\bibitem[{Chawla {et~al.}(2002)Chawla, Bowyer, Hall, \&
  Kegelmeyer}]{Chawla2002}
Chawla, N.~V., Bowyer, K.~W., Hall, L.~O., \& Kegelmeyer, W.~P. 2002, Journal
  of Artificial Intelligence Research, 16, 321

\bibitem[{Colless {et~al.}(2001)Colless, Dalton, Maddox, Sutherland, Norberg,
  Cole, Bland-Hawthorn, Bridges, Cannon, Collins, Couch, Cross, Deeley,
  DePropris, Driver, Efstathiou, Ellis, Frenk, Glazebrook, Jackson, Lahav,
  Lewis, Lumsden, Madgwick, Peacock, Peterson, Price, Seaborne, \&
  Taylor}]{Colless2001}
Colless, M., Dalton, G.~B., Maddox, S.~J., {et~al.} 2001, Monthly Notices of
  the Royal Astronomical Society, Volume 328, Issue 4, pp. 1039-1063., 328,
  1039

\bibitem[{Condon {et~al.}(1998)Condon, Cotton, Greisen, Yin, Perley, Taylor, \&
  Broderick}]{Condon1998}
Condon, J.~J., Cotton, W.~D., Greisen, E.~W., {et~al.} 1998, The Astronomical
  Journal, 115, 1693

\bibitem[{Cutri \& al.(2013)}]{Cutri2013}
Cutri, R., \& al., E. 2013, VizieR Online Data Catalog, 2328, 0

\bibitem[{D'Abrusco {et~al.}(2014)D'Abrusco, Massaro, Paggi, Smith, Masetti,
  Landoni, \& Tosti}]{DAbrusco2014}
D'Abrusco, R., Massaro, F., Paggi, A., {et~al.} 2014, The Astrophysical Journal
  Supplement Series, 215, 14

\bibitem[{Dom{\'{i}}nguez \& Ajello(2015)}]{Dominguez2015}
Dom{\'{i}}nguez, A., \& Ajello, M. 2015, The Astrophysical Journal Letters,
  Volume 813, Issue 2, article id. L34, 4 pp. (2015)., 813, arXiv:1510.07913

\bibitem[{Donato {et~al.}(2001)Donato, Ghisellini, Tagliaferri, \&
  Fossati}]{Donato2001}
Donato, D., Ghisellini, G., Tagliaferri, G., \& Fossati, G. 2001, Astronomy
  {\&} Astrophysics, 375, 739

\bibitem[{Evans {et~al.}(2007)Evans, Tyler, Beardmore, \& Osborne}]{Evans2007}
Evans, P.~A., Tyler, L.~G., Beardmore, A.~P., \& Osborne, J.~P. 2007

\bibitem[{Evans {et~al.}(2008)Evans, Beardmore, Page, Osborne, O'Brien,
  Willingale, Starling, Burrows, Godet, Vetere, Racusin, Goad, Wiersema,
  Angelini, Capalbi, Chincarini, Gehrels, Kennea, Margutti, Morris, Mountford,
  Pagani, Perri, Romano, \& Tanvir}]{Evans2008}
Evans, P.~A., Beardmore, A.~P., Page, K.~L., {et~al.} 2008, Monthly Notices of
  the Royal Astronomical Society, Volume 397, Issue 3, pp. 1177-1201., 397,
  1177

\bibitem[{Evans {et~al.}(2014)Evans, Osborne, Beardmore, Page, Willingale,
  Mountford, Pagani, Burrows, Kennea, Perri, Tagliaferri, \&
  Gehrels}]{Evans2014}
Evans, P.~A., Osborne, J.~P., Beardmore, A.~P., {et~al.} 2014, Astrophysical
  Journal, Supplement Series, 210, arXiv:1311.5368

\bibitem[{Fichtel {et~al.}(1993)Fichtel, Bertsch, Dingus, Hartman, Hunter,
  Kanbach, Kniffen, Kwok, Lin, Mattox, Mayer-Hasselwander, Michelson, von
  Montigny, Nolan, Pinkau, Rothermel, Schneid, Sommer, Sreekumar, \&
  Thompson}]{Fichtel1993}
Fichtel, C.~E., Bertsch, D.~L., Dingus, B., {et~al.} 1993, Advances in Space
  Research, 13, 637

\bibitem[{Flesch \& W.(2015)}]{Flesch2015}
Flesch, E.~W., \& W., E. 2015, Publications of the Astronomical Society of
  Australia, Volume 32, id.e010 17 pp., 32, arXiv:1502.06303

\bibitem[{Flesch \& W.(2017)}]{Flesch2017}
---. 2017, VizieR On-line Data Catalog: VII/277. Originally published in:
  2015PASA...32...10F, 7277

\bibitem[{Gehrels {et~al.}(2004)Gehrels, Chincarini, Giommi, Mason, Nousek,
  Wells, White, Barthelmy, Burrows, Cominsky, Hurley, Marshall, Meszaros,
  Roming, Angelini, Barbier, Belloni, Campana, Caraveo, Chester, Citterio,
  Cline, Cropper, Cummings, Dean, Feigelson, Fenimore, Frail, Fruchter,
  Garmire, Gendreau, Ghisellini, Greiner, Hill, Hunsberger, Krimm, Kulkarni,
  Kumar, Lebrun, Lloyd‐Ronning, Markwardt, Mattson, Mushotzky, Norris,
  Osborne, Paczynski, Palmer, Park, Parsons, Paul, Rees, Reynolds, Rhoads,
  Sasseen, Schaefer, Short, Smale, Smith, Stella, Tagliaferri, Takahashi,
  Tashiro, Townsley, Tueller, Turner, Vietri, Voges, Ward, Willingale, Zerbi,
  \& Zhang}]{Gehrels2004}
Gehrels, N., Chincarini, G., Giommi, P., {et~al.} 2004, The Astrophysical
  Journal, 611, 1005

\bibitem[{Hassan {et~al.}(2017)Hassan, Dom{\'{i}}nguez, Lefaucheur, Mazin,
  Pita, \& Zech}]{Hassan2017}
Hassan, T., Dom{\'{i}}nguez, A., Lefaucheur, J., {et~al.} 2017, eprint
  arXiv:1708.07704, arXiv:1708.07704

\bibitem[{Hastie {et~al.}(2009)Hastie, Tibshirani, \& Friedman}]{Hastie2009}
Hastie, T., Tibshirani, R., \& Friedman, J. 2009, {Springer Series in
  Statistics}, doi:10.1007/978-0-387-98135-2

\bibitem[{Hearst {et~al.}(1998)Hearst, Dumais, Osman, Platt, \&
  Scholkopf}]{Hearst1998}
Hearst, M.~a., Dumais, S.~T., Osman, E., Platt, J., \& Scholkopf, B. 1998,
  {\ldots}Systems and their {\ldots}, arXiv:arXiv:1011.1669v3

\bibitem[{Jones {et~al.}(2009)Jones, Read, Saunders, Colless, Jarrett, Parker,
  Fairall, Mauch, Sadler, Watson, Burton, Campbell, Cass, Croom, Dawe, Fiegert,
  Frankcombe, Hartley, Huchra, James, Kirby, Lahav, Lucey, Mamon, Moore,
  Peterson, Prior, Proust, Russell, Safouris, Wakamatsu, Westra, \&
  Williams}]{Jones2009}
Jones, D.~H., Read, M.~A., Saunders, W., {et~al.} 2009, Monthly Notices of the
  Royal Astronomical Society, Volume 399, Issue 2, pp. 683-698., 399, 683

\bibitem[{Kalberla {et~al.}(2005)Kalberla, Burton, Hartmann, Arnal, Bajaja,
  Morras, \& Poppel}]{Kalberla2005}
Kalberla, P. M.~W., Burton, W.~B., Hartmann, D., {et~al.} 2005, Astronomy and
  Astrophysics, Volume 440, Issue 2, September III 2005, pp.775-782, 440, 775

\bibitem[{Marchesi {et~al.}(2018)Marchesi, Kaur, \& Ajello}]{Marchesi2018}
Marchesi, S., Kaur, A., \& Ajello, M. 2018, The Astronomical Journal, 156, 212

\bibitem[{Massaro {et~al.}(2016)Massaro, Giommi, Leto, Marchegiani, Maselli,
  Perri, Piranomonte, \& Sclavi}]{Massaro2016}
Massaro, E., Giommi, P., Leto, C., {et~al.} 2016, VizieR On-line Data Catalog:
  VII/274. Originally published in: 2015Ap{\&}SS.357...75M, 7274

\bibitem[{Massaro {et~al.}(2015)Massaro, Maselli, Leto, Marchegiani, Perri,
  Giommi, \& Piranomonte}]{Massaro2015a}
Massaro, E., Maselli, A., Leto, C., {et~al.} 2015, Astrophysics and Space
  Science, Volume 357, Issue 1, article id.75 4pp., 357, arXiv:1502.07755

\bibitem[{Massaro {et~al.}(2012)Massaro, D'Abrusco, Tosti, Ajello, Gasparrini,
  Grindlay, \& Smith}]{Massaro2012}
Massaro, F., D'Abrusco, R., Tosti, G., {et~al.} 2012, Astrophysical Journal,
  750, 138

\bibitem[{Mauch {et~al.}(2003)Mauch, Murphy, Buttery, Curran, Hunstead,
  Piestrzynski, Robertson, \& Sadler}]{Mauch2003}
Mauch, T., Murphy, T., Buttery, H.~J., {et~al.} 2003, Monthly Notice of the
  Royal Astronomical Society, Volume 342, Issue 4, pp. 1117-1130., 342, 1117

\bibitem[{Mirabal {et~al.}(2016)Mirabal, Charles, Ferrara, Gonthier, Harding,
  S{\'{a}}nchez-Conde, \& Thompson}]{Mirabal2016}
Mirabal, N., Charles, E., Ferrara, E.~C., {et~al.} 2016, The Astrophysical
  Journal, 825, 69

\bibitem[{Mirabal {et~al.}(2012)Mirabal, Fr{\'{i}}as-Martinez, Hassan, \&
  Fr{\'{i}}as-Martinez}]{Mirabal2012}
Mirabal, N., Fr{\'{i}}as-Martinez, V., Hassan, T., \& Fr{\'{i}}as-Martinez, E.
  2012, Monthly Notices of the Royal Astronomical Society: Letters, 424, L64

\bibitem[{Paiano {et~al.}(2017)Paiano, Franceschini, \& Stamerra}]{Paiano2017}
Paiano, S., Franceschini, A., \& Stamerra, A. 2017, Monthly Notices of the
  Royal Astronomical Society, Volume 468, Issue 4, p.4902-4937, 468, 4902

\bibitem[{Paturel {et~al.}(2003)Paturel, Petit, Prugniel, Theureau, Rousseau,
  Brouty, Dubois, \& Cambr?sy}]{Paturel2003}
Paturel, G., Petit, C., Prugniel, P., {et~al.} 2003, Astronomy {\{}{\&}{\}}
  Astrophysics, 412, 45

\bibitem[{Pedregosa {et~al.}(2011)Pedregosa, Weiss, \& Brucher}]{Pedregosa2011}
Pedregosa, F., Weiss, R., \& Brucher, M. 2011, Journal of Machine Learning
  Research, arXiv:1201.0490v1

\bibitem[{Quinlan \& Shapiro(1990)}]{Quinlan1990}
Quinlan, G.~D., \& Shapiro, S.~L. 1990, The Astrophysical Journal, 356, 483

\bibitem[{Salvetti {et~al.}(2017)Salvetti, Chiaro, {La Mura}, \&
  Thompson}]{Salvetti2017}
Salvetti, D., Chiaro, G., {La Mura}, G., \& Thompson, D.~J. 2017, Monthly
  Notices of the Royal Astronomical Society, Volume 470, Issue 2, p.1291-1297,
  470, 1291

\bibitem[{{Saz Parkinson} {et~al.}(2016){Saz Parkinson}, Xu, Yu, Salvetti,
  Marelli, \& Falcone}]{Parkinson2016}
{Saz Parkinson}, P.~M., Xu, H., Yu, P. L.~H., {et~al.} 2016, The Astrophysical
  Journal, 820, 8

\bibitem[{Sharma \& Chauhan(2011)}]{Wright2010}
Sharma, S.~K., \& Chauhan, R. 2011, Current Science, 101, 308

\bibitem[{Stroh \& Falcone(2013)}]{Stroh2013}
Stroh, M.~C., \& Falcone, A.~D. 2013, The Astrophysical Journal Supplement,
  Volume 207, Issue 2, article id. 28, 12 pp. (2013)., 207,
  doi:10.1088/0067-0049/207/2/28

\bibitem[{Taylor(2005)}]{Taylor2005}
Taylor, M.~B. 2005, in Astronomical Data Analysis Software and Systems XIV,
  Vol. 347, 29--+

\bibitem[{Voges {et~al.}(1999)Voges, Aschenbach, Boller, Braeuninger, Briel,
  Burkert, Dennerl, Englhauser, Gruber, Haberl, Hartner, Hasinger, Pfeffermann,
  Pietsch, Predehl, Rosso, Schmitt, Truemper, Zimmermann, \&
  Zimmermann}]{Voges1999}
Voges, W., Aschenbach, B., Boller, T., {et~al.} 1999, Astronomy and
  Astrophysics, v.349, p.389-405 (1999), 349, 389

\bibitem[{Voges {et~al.}(2000)Voges, Aschenbach, Boller, Brauninger, Briel,
  Burkert, Dennerl, Englhauser, Gruber, Haberl, Hartner, Hasinger, Pfeffermann,
  Pietsch, Predehl, Schmitt, Trumper, \& Zimmermann}]{Voges2000}
---. 2000, International Astronomical Union Circular, 7432, 1

\bibitem[{Wenger {et~al.}(2000)Wenger, Ochsenbein, Egret, Dubois, Bonnarel,
  Borde, Genova, Jasniewicz, Laloe, Lesteven, \& Monier}]{Wenger2000}
Wenger, M., Ochsenbein, F., Egret, D., {et~al.} 2000, Astronomy and
  Astrophysics Supplement, v.143, p.9-22, 143, 9

\end{thebibliography}
\end{document}